\documentclass[12pt,a4paper]{article}
\usepackage[T1]{fontenc} 
\usepackage{ae,aecompl}  
\usepackage{amsmath}
\usepackage{amssymb}
\usepackage{booktabs}  
\usepackage{longtable} 
\usepackage{lscape}    
\usepackage{xspace} 
\usepackage[usenames]{color} 
\usepackage[flushmargin,hang]{footmisc} 
\usepackage{cite} 
\usepackage{array}
\usepackage{a4wide}
\usepackage{epsfig}
\usepackage{caption}
\usepackage{subfig}
\usepackage{psfrag} 

\psfrag{meff}{{\scriptsize $\langle m_\nu \rangle$ [eV]}}
\psfrag{s12}{{\scriptsize $\sin^2 \theta_{12}$}}
\psfrag{T12}{{\scriptsize $T_{1/2}^{ \, 0 \nu } $ [10$^{27}$ y]}}
\psfrag{dma}{{\scriptsize $\Delta m_{\rm A}^2$ [10$^{-3}$ eV$^2$]}}
\psfrag{NME}{{$M^{\prime 0\nu}$}}
\psfrag{mihmax}{{\scriptsize $\langle m_\nu \rangle_{\rm max}^{\rm IH}$}}

\definecolor{MyDarkGreen}{rgb}{0,0.63,0.23} 
\definecolor{MyDarkBlue}{rgb}{0.23,0.21,0.69} 
\definecolor{MyLightBlue}{rgb}{0.22,0.51,0.86}

\usepackage[colorlinks=true,linkcolor=MyDarkBlue,citecolor=MyDarkGreen,urlcolor=MyLightBlue,bookmarksnumbered=true,bookmarksopen]{hyperref} 

\newdimen\LTcapwidth \LTcapwidth=\textwidth
\newcommand{\nubb}{$0\nu\beta\beta$\xspace}
\newcommand{\nubbd}{$0\nu\beta\beta$ decay\xspace}
\def\be{\begin{equation}}
\def\ee{\end{equation}}
\newcommand{\ba}{\begin{array}{c}}
\newcommand{\bad}{\begin{array}{ccc}}
\newcommand{\ea}{\end{array}}
\newcommand{\meff}{\langle m_\nu \rangle}

\newcommand{\dms}{\mbox{$\Delta m^2_{\odot}$}}
\newcommand{\dma}{\mbox{$\Delta m^2_{\rm A}$}}

\def\gsim{\mathrel{
   \rlap{\raise 0.511ex \hbox{$>$}}{\lower 0.511ex \hbox{$\sim$}}}}
\def\lsim{\mathrel{
   \rlap{\raise 0.511ex \hbox{$<$}}{\lower 0.511ex \hbox{$\sim$}}}}

\begin{document}

\title{\vspace{-2cm}
\hfill {\small }\\[-0.1in]
\vskip 0.8cm
\bf \Large
Neutrinoless Double Beta Decay, the Inverted Hierarchy and 
Precision Determination of $\theta_{12}$ 
}
\author{
~~Alexander Dueck$^a$\thanks{email: 
\tt alexander.dueck@mpi-hd.mpg.de}\mbox{ },~~
Werner Rodejohann$^a$\thanks{email: 
\tt werner.rodejohann@mpi-hd.mpg.de}\mbox{
},~~
Kai Zuber$^b$\thanks{email: 
\tt zuber@physik.tu-dresden.de}
\\\\
{\normalsize \it$^a$Max--Planck--Institut f\"ur Kernphysik,}\\
{\normalsize \it  Postfach 103980, D--69029 Heidelberg, Germany}  \\
\\ 
{\normalsize \it $^b$Technische Universit\"at Dresden,}\\
{\normalsize \it Institut f\"ur Kern-- und Teilchenphysik,}\\
{\normalsize \it D--01069 Dresden, Germany}\\
} 
\date{}
\maketitle
\thispagestyle{empty}
\vspace{0.18cm}
\begin{abstract}
\noindent  
Ruling out the inverted neutrino hierarchy with neutrinoless double
beta decay experiments is possible if a limit on the effective mass below the
minimal theoretically possible value is reached.  
We stress that this lower limit depends strongly 
on the value of the solar neutrino mixing angle: it introduces an
uncertainty of a factor of 
2 within its current 3$\sigma$ range. 
If an experiment is not background-free, a factor of two in effective mass corresponds 
to a combined factor of 16 improvement for the  
experimental parameters running time, detector mass, background level and 
energy resolution. Therefore, a more precise determination of
$\theta_{12}$ is crucial for the interpretation of 
experimental results and the evaluation of the potential and requirements for future experiments. 
We give the required half-lifes to exclude (and touch) the inverted
hierarchy regime for all double beta decay isotopes with a $Q$-value
above 2 MeV. The nuclear matrix elements 
from 6 different groups and, if available, their errors are used and
compared. We carefully put the calculations on 
equal footing in what regards various convention issues.  
We also use our compilation of matrix elements to give the reachable values of the
effective mass for a given half-life value. 

\end{abstract}

\newpage

\section{Introduction}
Neutrinoless Double Beta Decay (\nubb) is a process of fundamental
importance for particle physics
\cite{tomoda91,Vergados:2002pv,Avignone:2007fu}. 
In the best motivated interpretation \cite{WR}
of this process, light Majorana neutrinos, whose mixing is observed in
neutrino oscillation experiments, 
are exchanged in the process, and the particle physics quantity which is probed is the "effective mass"
\be
\langle m_\nu \rangle
= \left| U_{e1}^2 \, m_1 + U_{e2}^2 \, m_2 \, e^{i \alpha} 
+ U_{e3}^2 \, m_3 \, e^{i\beta}\right| . 
\ee  
Here $U_{e1} = \cos \theta_{12} \, \cos \theta_{13}$, $U_{e2} = \sin
\theta_{12} \, \cos \theta_{13}$ and $U_{e3}^2 = 1 - U_{e1} ^2-
U_{e2}^2$. The current knowledge of  
these mixing angles is given in Table \ref{tab:params}. The lifetime
of \nubbd is inversely proportional 
to the effective mass squared. 

Apart from verifying the Majorana nature of neutrinos, the effective
mass depends on a number of known and unknown neutrino parameters, and
testing or cross-checking the values of these parameters is obviously
an immensely important task. Among the unknown neutrino
parameters the neutrino mass ordering (the sign of the atmospheric
mass-squared difference) is of particular interest. 
It is indeed an exciting possibility to rule out the inverted ordering
(IH) with \nubb. This is possible because the lower limit of the
effective mass is non-zero in this case \cite{petcov,Petcov:2005yq}. 
Actually, if at the time when the inverted hierarchy regime is under
test at double beta decay experiments the 
mass ordering is known to be inverted (by an oscillation experiment or
by a galactic supernova explosion), 
then testing the inverted hierarchy means testing directly the
Majorana nature of neutrinos. If the mass ordering is not known, the
experiments can rule out the inverted hierarchy only if in addition
the Majorana nature of neutrinos is assumed. However, this happens  in
the vast majority of models and scenarios leading to neutrino mass,
and is also natural from an effective field theory point of view. 

In any case, a natural scale for the effective mass provided by
particle physics is the minimal value of the effective mass in the
inverted hierarchy, and should be the 
intermediate- or long-term aim of double beta experiments. 

We stress in this paper that the lower limit of the effective mass is
a sensitive function of the solar neutrino mixing angle $\theta_{12}$: the current 
$3\sigma $ range of $\theta_{12}$ introduces an uncertainty of a
factor of 2. In realistic, i.e.~background dominated, experiments the
achievable half-life reach is  proportional to 
\be
T_{1/2}^{0\nu} \propto a \times \epsilon \times \sqrt{\frac{M \times t}{B \times
\Delta E}},
\label{eqn:hlsensbg}
\ee
where $a$ is the isotopical
abundance of the double beta emitter,
$M$ the used mass, $t$ the measuring time,
$\epsilon$ the detection efficiency, $\Delta E$ the energy
resolution at the peak position and $B$ the background index typically given in
counts/keV/kg/yr.
\begin{table}[t]
\centering
\begin{tabular}{@{}ccc@{}}
\toprule
Parameter  & $\mbox{Best-fit}^{+1 \sigma}_
    {-1 \sigma}$ & $3 \sigma$ \\ \midrule
$\sin^2 \theta_{12}$ & 
    0.318$^{+0.019}_{-0.016}$  & 0.27-0.38 \\ \addlinespace
$\sin^2 \theta_{13}$ & 
    0.013$^{+0.013}_{-0.009}$  & $\le 0.053$ \\ \addlinespace
$\Delta m_{\rm A}^2$ [10$^{-3}$ eV$^2$] & 2.40$^{+0.12}_{-0.11}$ & 2.07-2.75 \\ \addlinespace
$\Delta m_\odot^2$ [10$^{-5}$ eV$^2$] & 7.59$^{+0.23}_{-0.18}$ & 7.03-8.27 \\
\bottomrule
\end{tabular}
\caption{Neutrino mixing parameters: best-fit values as well
as $1\sigma$ and 3$\sigma$ ranges \cite{schwetz2010:fit}.}
\label{tab:params}
\end{table}
Hence, an uncertainty of 2 in the effective mass corresponds to a factor of 
$2^2 = 4$ in terms of lifetime reach and a factor of 
 $2^4 = 16$ uncertainty in the above combination of experimental parameters. 
In this work we aim to stress this fact and to illustrate its
consequences. We quantify the
requirements to test the inverted hierarchy in terms of 
necessary half-life reach. We consider all 
\nubb-isotopes with a $Q$-value above 2 MeV and compile the nuclear matrix
element calculations from six different groups. That is, we study the
isotopes $^{48}$Ca, $^{76}$Ge, $^{82}$Se, $^{96}$Zr, $^{100}$Mo, $^{110}$Pd, $^{116}$Cd,
$^{124}$Sn, $^{130}$Te, $^{136}$Xe, and $^{150}$Nd, as well as nuclear
matrix element calculations applying QRPA \cite{faessler:dblbeta,suhonen1998}, 
Nuclear Shell Model \cite{caurier2009}, 
the Interacting Boson Model \cite{ibm:2009}, the Generating Coordinate
Method \cite{Rodriguez:2010mn}, and the
projected-Hartree-Fock-Bogoliubov model \cite{PhysRevC.82.064310}. 
Particular care is taken to put the calculations on equal footing in
what regards various convention issues, such as the axial vector
coupling $g_A$ and the nuclear radius appearing in the phase space factor. 
 We present the results for different values of $\theta_{12}$, in order to
show its impact. 

We are taking the point of view that the spread of
nuclear matrix elements and lifetimes obtained in our analysis is a
fair estimate of the true allowed range. Though experimental approaches to
reduce the uncertainty \cite{Zuber:2005fu}, and statistical approaches
to better estimate the theoretical uncertainties (see e.g.~\cite{Faessler:2008xj}), have
started, at the current stage the collection of available results and
the use of their spread is the most pragmatic procedure.\\

Nevertheless, our main conclusions are independent of this and 
quite straightforward: a precision
determination of the solar neutrino mixing angle 
is crucial to determine the physics potential of, and
requirements for, neutrinoless double beta decay experiments. 
Some proposals for solar neutrino experiments which can pin down 
$\theta_{12}$ more precisely can be found in the literature 
\cite{SNO+,CLEAN,HERON,LENS}. Large-scale long baseline reactor neutrino experiments
have also been proposed \cite{Bandyopadhyay:2003du,Minakata:2004jt,Bandyopadhyay:2004cp,Petcov:2006gy}, but
to our knowledge still await detailed study by experimentalists. 
The main focus of future precision neutrino oscillation physics is put
on mass ordering, the other mixing angles and CP violation in facilities such
as super-, beta-beams or neutrino factories. 
Given the impact of $\theta_{12}$ on neutrinoless double beta decay that we discuss
here, we hope to provide additional motivation for studies and
proposals in order to determine $\theta_{12}$ as precisely as
possible\footnote{The additional physics potential of precision solar neutrino or
$\theta_{12}$ experiments is e.g.~solving the metalicity
problem of the Sun \cite{metal}, probing the transition region of the
electron neutrino survival probability in the Sun's interior
\cite{transition}, or distinguishing theoretical approaches to lepton
mixing such as tri-bimaximal mixing from alternative models
\cite{ADR}.}. At least we encourage to seriously determine and
optimize the potential
of future experiments in what regards the achievable precision of
$\theta_{12}$.

Using our compilation of matrix element calculations, we also present
results for the necessary half-life in order to touch the inverted
hierarchy regime. Finally, we investigate which limits on the 
effective mass can be achieved for a given half-life, and what the
current limits are. These points are independent of the value of
$\theta_{12}$.

We find that the isotope $^{100}$Mo tends to be interesting, 
in the sense that with the same lifetime it  
can slightly more easily rule out the inverted hierarchy, or achieve the better
limit on the effective mass. This may be helpful for experiments considering
various alternative isotopes to study. \\ 

The paper is built up as follows: in Section \ref{sec:meff} we shortly
discuss the effective mass and its dependence on the solar neutrino
mixing angle in the inverted hierarchy. Section \ref{sec:hl} deals
with the various calculations of the nuclear matrix elements and their
impact on ruling out and touching the inverted hierarchy regime. We
point out the difficulties arising from the chosen convention, 
which can arise by comparing different 
nuclear matrix element calculations. 
The general limits on the effective mass as a function of an
achievable half-life are given in a short Section 
\ref{sec:fut}, where also current limits on the effective mass are
compiled. In Section \ref{sec:exp} we give some examples on the
experimental consequences of our results for future experiments.  
Tables and details are delegated to the Appendices, and
conclusions are presented in Section \ref{sec:concl}.

\section{Effective neutrino mass and experimental values of neutrino oscillation parameters\label{sec:meff}}

\begin{table}[t]
\centering
\begin{tabular}{@{}l@{\hspace*{0.4cm}}cc@{}}
\toprule
 & \multicolumn{2}{c}{$\langle m _\nu\rangle^{\text{IH}}_{\text{min}}$ [eV]} \\ \cmidrule{2-3}
$\sin^2 \theta_{12}$ & minimal & maximal \\ \midrule
0.270 & 0.0196 & 0.0240 \\
0.318 & 0.0154 & 0.0189 \\
0.380 & 0.0100 & 0.0123 \\
\bottomrule
\end{tabular}
\caption{Lower limit of the effective electron neutrino mass in
the case of an inverted hierarchy for
different values of $\sin^2 \theta_{12}$. The minimal and maximal
values are obtained by varying $\Delta m_{\rm A}^2$, $\Delta
m_\odot^2$ and $\sin^2 \theta_{13}$ in their allowed 3$\sigma$ ranges.}
\label{tab:meff}
\end{table}

In general, the decay rate of \nubbd factorizes in a kinematical, nuclear
physics and particle physics part: 
\be \label{eq:Ggen}
\Gamma^{0\nu} = G_{\rm kin} \, |M_{\rm nucl}|^2 \, X_{\rm part}
\, . 
\ee
The observation of the decay would establish the nature of the
neutrino as a Majorana particle \cite{Schechter82}, independent on
whether indeed light Majorana neutrinos are exchanged in the diagram
leading to \nubb. However, the most natural interpretation is indeed
that this is the case, because we know that neutrinos have a non-vanishing
rest mass, and
in the vast majority of models they are Majorana particles. The
particle physics parameter in the decay width Eq.~(\ref{eq:Ggen}) is 
therefore $X_{\rm part} \propto \meff^2$, where $\meff$ is the effective electron neutrino
mass defined as 
\begin{equation}
\label{eqn:meff}
\langle m_\nu \rangle = \left| 
 c_{12}^2 \, c_{13}^2  \, m_1 + s_{12}^2 \, c_{13}^2  \,  m_2 \, e^{i \alpha} 
+ s_{13}^2 \, m_3 \, e^{i\beta}\right| , 
\end{equation}
where $c_{ij} = \cos \theta_{ij},\ s_{ij} = \sin \theta_{ij}$, and
$\alpha$, $\beta$ are the two Majorana phases. 
It depends on the three neutrino mass eigenstates $m_i$ and the first
row of the Pontecorvo-Maki-Nakagawa-Sakata (PMNS) mixing matrix. 
The effective mass, $\langle m_\nu \rangle$, can span a wide range due
to the unknown Majorana phases, the unknown total neutrino mass scale,
and the unknown mass ordering. 
We are interested here mostly in the case of the inverted hierarchy
(IH), which corresponds to $m_2 > m_1 > m_3$. In this case the maximum
and minimum values of $\langle m_\nu \rangle$ are given by (see e.g.,
\cite{petcov,Petcov:2005yq,lindner:t13mnu}) 
\begin{equation}
\label{eqn:meffmaxih0}
\langle m_\nu \rangle^{\text{IH}}_{\text{max}} = \sqrt{m_3^2 + \Delta
m_{\rm A}^2} \, c_{12}^2 c_{13}^2 + 
\sqrt{m_3^2 + \Delta m^2_{\odot} + \Delta m_{\rm A}^2} \, s_{12}^2 c_{13}^2 +
m_3 s_{13}^2 \, ,
\end{equation}
and
\begin{equation}
\langle m_\nu \rangle^{\text{IH}}_{\text{min}} = \sqrt{m_3^2 + \Delta
m_{\rm A}^2} \, c_{12}^2 c_{13}^2 - 
\sqrt{m_3^2 + \Delta m^2_{\odot} + \Delta m_{\rm A}^2} \, s_{12}^2 c_{13}^2 -
m_3 s_{13}^2 \, ,
\label{eqn:meffminih0}
\end{equation}
respectively. Here, $\Delta m_\odot^2 = m_2^2 - m_1^2$
is the solar and $\Delta m_{\rm A}^2 = |m_3^2 - m_1^2|$ 
the atmospheric mass-squared difference. The values we use for the
mixing parameters are shown in Table \ref{tab:params}.

\begin{figure}[t]
\centering
\subfloat[]{\label{fig:meff-t12}\epsfig{file=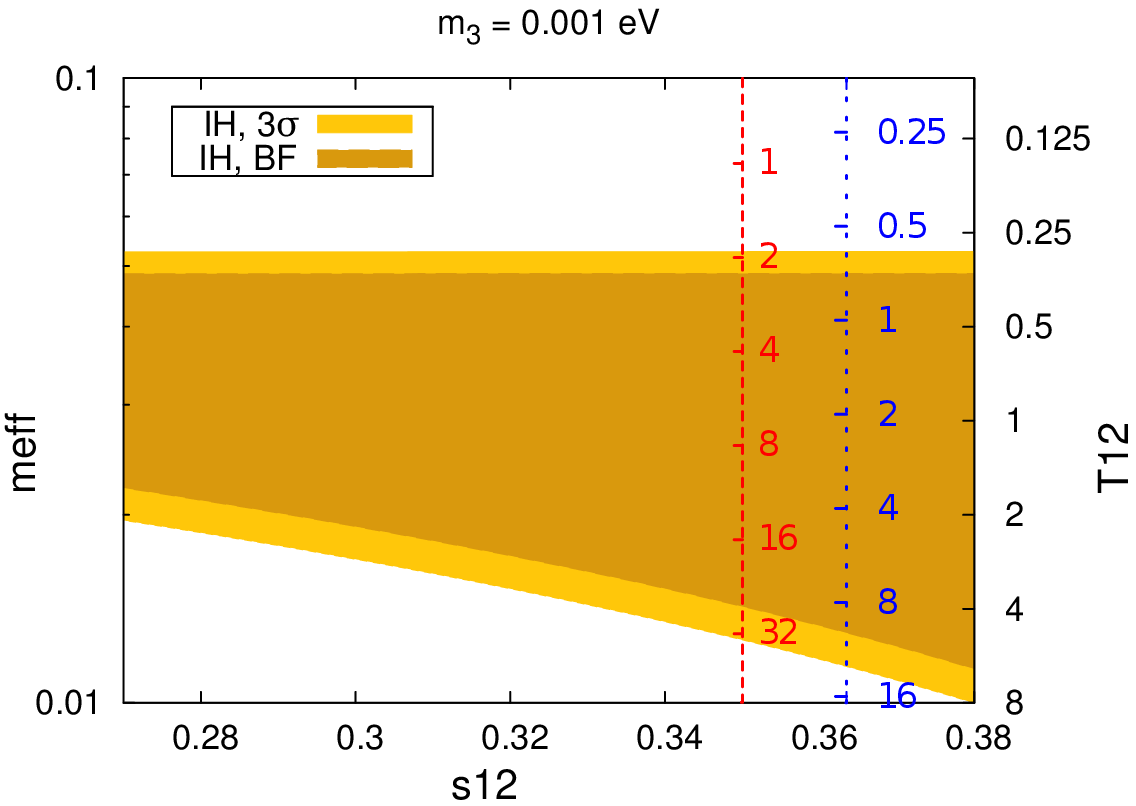,width=7cm,height=5cm}}\hspace{.85cm}
\subfloat[]{\label{fig:meff-dma}\epsfig{file=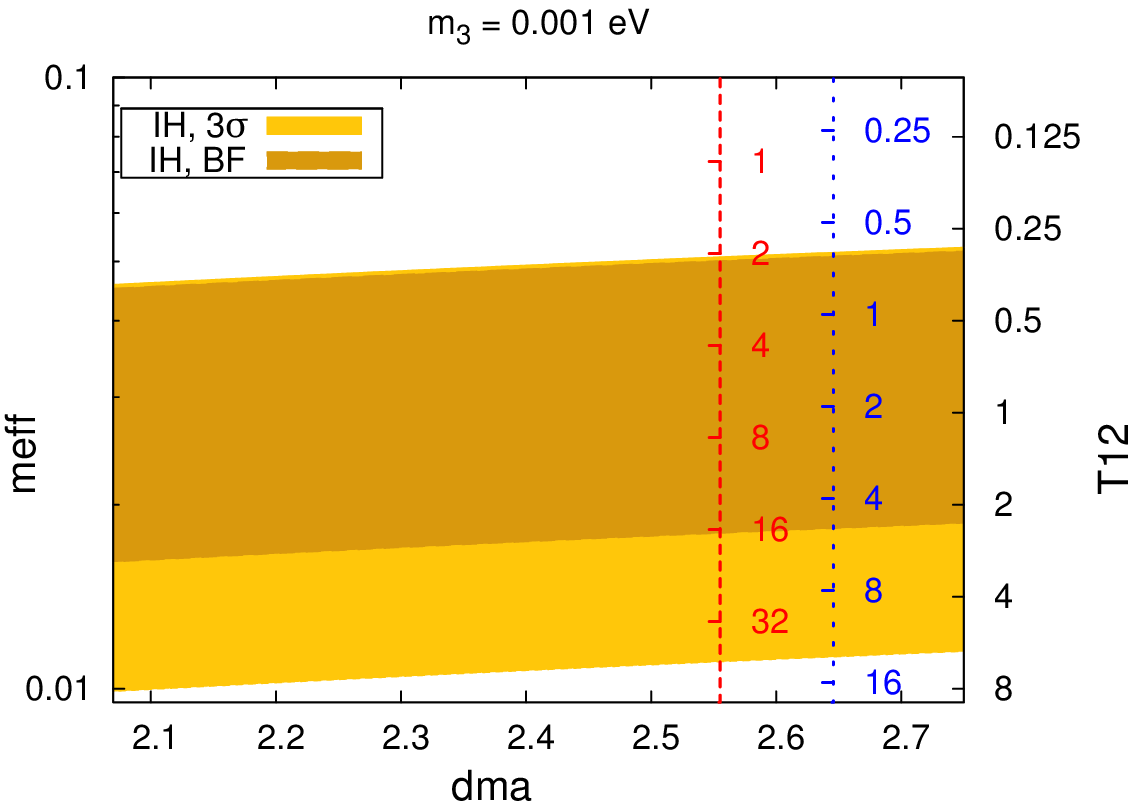,width=7cm,height=5cm}}

\caption{The effective electron neutrino mass in the case of an
inverted hierarchy is shown as a 
function of \subref{fig:meff-t12} $\sin^2\theta_{12}$ and 
\subref{fig:meff-dma} $\Delta m_{\rm A}^2$ with best-fit values and
3$\sigma$ ranges for the other oscillation parameters. On the right
side of the plots the corresponding half-life for $^{76}$Ge is shown
assuming three different nuclear matrix elements: $M^{\prime0\nu}=2.81$ (red dashed
axis), $M^{\prime0\nu}=5$ (blue dotted axis), 
and $M^{\prime0\nu}=7.24$ (black solid axis).}
\label{3figs}
\end{figure}

Unless the smallest mass $m_3$ is larger than about $0.05$ eV, the effective
mass does basically not depend on its value, and increases linearly with $m_3$ afterwards. 
In the case of $m_3 \lsim 0.05$ eV, one finds 
\begin{equation}
\label{eqn:meffmaxih}
\langle m_\nu \rangle^{\text{IH}}_{\text{max}} \simeq c_{13}^2 \,
\sqrt{\dma} \, , 
\end{equation}
and
\begin{equation}
\langle m_\nu \rangle^{\text{IH}}_{\text{min}} \simeq c_{13}^2 \,
\sqrt{\dma} \, \cos 2 \theta_{12}  = 
\left(1 - |U_{e3}|^2 \right)  \sqrt{\dma} \left(1 - 2 \, \sin^2
\theta_{12} \right) , 
\label{eqn:meffminih}
\end{equation}
respectively. The maximal value is obtained for $\alpha = 0$ and the
minimal value for $\alpha = \pi/2$. Since $\theta_{12}$ is non-maximal,
the minimal value of $\meff$~is non-zero, which is in contrast to the
normal mass ordering, in which the effective mass can vanish. By
obtaining experimentally an upper limit on the effective mass below 
$\langle m_\nu \rangle^{\text{IH}}_{\text{min}}$, we can rule out the
inverted ordering. If we would know by independent evidence that the
ordering is inverted (i.e., from a long baseline experiment, or
observation of a galactic supernova), then obtaining such an upper
limit would even mean that the Majorana nature of neutrinos would have
been ruled out.  

From a more pragmatic point of view, particle physics provides a scale
for limits on the effective mass, which should be the sensitivity goals of the 
experimental program. These values are $\langle m_\nu 
\rangle^{\text{IH}}_{\text{max}}$ and $\langle m_\nu
\rangle^{\text{IH}}_{\text{min}}$ given in Eq.~(\ref{eqn:meffmaxih}) and (\ref{eqn:meffminih}),
respectively. 
In Fig.~\ref{3figs} we show the effective mass for the
best-fit and the $3\sigma$ ranges of the oscillation parameters as a
function of $\sin^2 \theta_{12}$ and \dma. It is clear that the
dependence of the lower limit on $\sin^2 \theta_{12}$ is very strong. 
In the currently allowed $3\sigma$ range the range of
$\theta_{12}$ quantifies to a factor of 2 uncertainty for $\langle m_\nu
\rangle^{\text{IH}}_{\text{min}}$, which
translates into a factor $2^2 = 4$ in lifetime reach for an
experiment. We illustrate this in the plots by translating $\meff$ into
the half-life for $^{76}$Ge for three representative values of the
nuclear matrix elements (see Section \ref{sec:NME}). 
Table \ref{tab:meff} shows the numerical values of the effective
electron neutrino mass in the case of an inverted hierarchy for
different values of the 
solar neutrino parameter $\sin^2 \theta_{12}$. 
The uncertainty in the other parameters $|U_{e3}|$ and \dma~is by far
not as significant, it amounts in total to a factor less than 25
\%. An extensive program to test \dma~and $|U_{e3}|$ is underway
(see e.g.~\cite{ISS}) and will have decreased this uncertainty considerably 
by the time the \nubb-experiments of the required sensitivity are running.  
The maximal value of the effective mass does not depend on
$\theta_{12}$, and hence its value is uncertain by less than 25 \%.

\section{Half-life sensitivities and the inverted hierarchy\label{sec:hl}}

We have seen above that in order to rule out the inverted ordering,
and to evaluate the physics potential of future experiments, the
value of $\theta_{12}$ is of crucial importance. 
We will now attempt to quantify the impact of $\theta_{12}$ in terms
of experimentally required half-life. Towards this end, we will have
to care with the available calculations of the nuclear matrix
elements (NMEs). We have scanned the literature and extracted the NME
values for five different calculational approaches of six different
groups. If given by the respective authors, we 
include the error estimates in the calculations for our results. 
In order to compare them in a proper way, we carefully try to put the
NMEs on equal footing, because details of conventions are often
different in different publications. We then consider all 11 potential
\nubb-isotopes with a $Q$-value above 2 MeV. We discuss the
necessary half-lifes to rule out and to touch the
inverted hierarchy, putting particular emphasis on the $\theta_{12}$-dependence
if necessary. Finally, using our compilation we also give the limits 
on $\meff$ as a function of future half-life limits for the 11
interesting isotopes. Using the
published half-life limits of different isotopes, we also 
give the current limits on the effective mass.

\subsection{\label{sec:NME}Nuclear Matrix Elements and the Half-life}

\begin{table}[t]
\centering
\begin{tabular}{@{}cccc@{}}
\toprule
Isotope		&	$G^{0\nu}$ [10$^{-14}$ yrs$^{-1}$]  &  $Q$ [keV]
& nat.~abund.~[\%]\\ \midrule
$^{48}$Ca	&	6.35    &  4273.7 & 0.187 \\
$^{76}$Ge	&	0.623   &  2039.1  & 7.8 \\
$^{82}$Se	&	2.70    &  2995.5 & 9.2 \\
$^{96}$Zr	&	5.63    &  3347.7 & 2.8 \\
$^{100}$Mo	&	4.36    &  3035.0   & 9.6 \\
$^{110}$Pd	&	1.40    &  2004.0 & 11.8 \\
$^{116}$Cd	&	4.62    &  2809.1 & 7.6 \\
$^{124}$Sn	&	2.55    &  2287.7 & 5.6 \\
$^{130}$Te	&	4.09    &  2530.3  & 34.5 \\
$^{136}$Xe	&	4.31    &  2461.9  & 8.9 \\
$^{150}$Nd	&	19.2    &  3367.3  & 5.6 \\ \bottomrule
\end{tabular}
\caption{$G^{0\nu}$ for different isotopes using $r_0=1.2$ fm. 
Values taken from Table 6 of Ref.~\cite{suhonen1998} 
($G^{0\nu}_1$ in their notation) and scaled to $g_A=1.25$ ($G^{0\nu}$
of $^{110}$Pd taken from Table IV of Ref.~\cite{PhysRevC.82.064310}). 
Also shown is the $Q$-value for the ground-state-to-ground-state transition which is calculated using 
isotope masses from Ref.~\cite{Audi2003337}
and the natural abundance in percent. Note that there is a misprint in Ref.~\cite{suhonen1998}, which
quotes $G^{0\nu}$ for $^{100}$Mo as $11.3 \times 10^{-14}$ yrs$^{-1}$.}
\label{tab:g0nu}
\end{table}

The \nubbd half-life is given according to Eq.~\eqref{eq:Ggen} by\footnote{Note that sometimes the
factor $1/m_e^{2}$ is carried into the definition of $G^{0\nu}$.} \cite{bilenky2010:0nbb}
\begin{equation}
(T^{0\nu}_{1/2})^{-1} = G^{0\nu} \left|M^{0\nu}\right|^2 
\left( \frac{\langle m_\nu \rangle}{m_e} \right)^2 ,
\label{eqn:half-life}
\end{equation}
where $G^{0\nu}$ is the phase space factor, $M^{0\nu}$ the NME, $m_e$
the electron mass, and the effective electron neutrino mass 
$\langle m_\nu \rangle$ as given in Eq.~\eqref{eqn:meff}. It is
known that the conversion of a lifetime into an effective mass, in
particular when different NMEs are compared, should be performed
carefully \cite{cowell2006:scaling,smolnikov:hl2meff}. 
The nuclear physics parameters, for instance the axial-vector coupling
$g_A$ lying in the range $1 \lsim g_A \lsim 1.25 $, 
should strictly speaking introduce an uncertainty in the value 
of $M^{0\nu}$ only. However, it is convention to include $g_A$ in the
phase space factor as well. In addition, the nuclear radius $R_A = r_0
A^{1/3}$ ($A$ being the atomic number) appears in 
$G^{0\nu}$, and there are differences in the normalization of $R_A$ with 
$r_0$, which should be taken into account. 
This leads to the small complication that NMEs calculated with
different values for $g_A$ and $r_0$ cannot be directly compared with
each other, since they have different phase space factors and hence
seemingly equal (by their value) matrix elements will lead to 
different decay half-lifes \cite{cowell2006:scaling} 
(see also the Appendix of \cite{tuebingen:g0nu}). We will outline
these issues in more detail in what follows.

\begin{figure}[tb]
\begin{center}
\epsfig{file=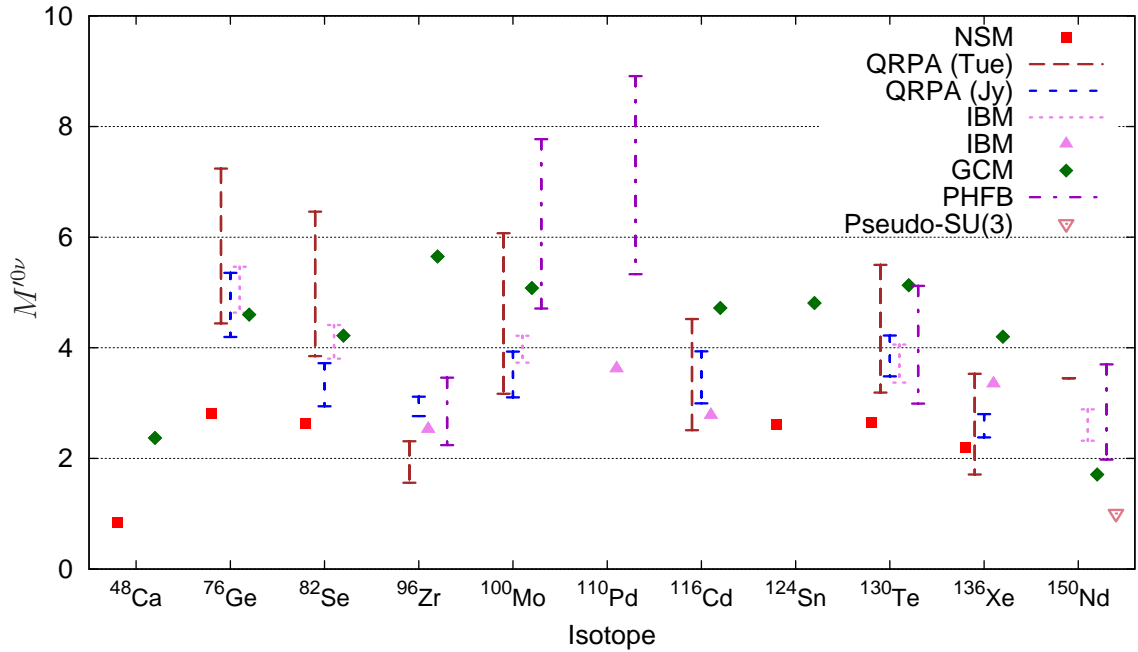}
\caption{\label{fig:isonme}NMEs calculated in different frameworks. 
We have scaled the cited values to $r_0 = 1.2$ fm and  $g_A = 1.25$
(see Eq.~\eqref{eqn:mp}) to make them directly comparable.
The exact values are given in Table \ref{tab:nme}.}
\end{center}
\end{figure}

\begin{table}[tb]
\small
\centering
\begin{tabular}{@{}lc@{~~ }c@{~~ }c@{~~ }c@{~~ }c@{~~ }c@{}}
\toprule
    &     NSM \cite{caurier2009}  &     T\"u
\cite{tuebingen2009,tuebingen:150nd} &  Jy  \cite{jyvaskyla2009} & IBM
\cite{ibm:2009} & GCM   \cite{Rodriguez:2010mn} &
PHFB\cite{PhysRevC.82.064310}  \\
Isotope    &     (UCOM)   &     (CCM)  &  (UCOM)     &  (Jastrow)    &  (UCOM) &  (mixed)  \\ \midrule
$^{48}$Ca  &        0.85       &       -       &       -         &       -         &      2.37      &       -      \\
$^{76}$Ge  &        2.81       &  4.44 - 7.24  &  4.195 - 5.355  &  4.636 - 5.465  &      4.6       &       -      \\
$^{82}$Se  &        2.64       &  3.85 - 6.46  &  2.942 - 3.722  &  3.805 - 4.412  &      4.22      &       -      \\
$^{96}$Zr  &          -        &  1.56 - 2.31  &  2.764 - 3.117  &      2.530      &      5.65      &  2.24 - 3.46 \\
$^{100}$Mo &          -        &  3.17 - 6.07  &  3.103 - 3.931  &  3.732 - 4.217  &      5.08      &  4.71 - 7.77 \\
$^{110}$Pd &          -        &       -       &       -         &      3.623      &       -        &  5.33 - 8.91 \\
$^{116}$Cd &          -        &  2.51 - 4.52  &  2.996 - 3.935  &      2.782      &      4.72      &       -      \\
$^{124}$Sn &        2.62       &       -       &       -         &       -         &      4.81      &       -      \\
$^{130}$Te &        2.65       &  3.19 - 5.50  &  3.483 - 4.221  &  3.372 - 4.059  &      5.13      &  2.99 - 5.12 \\
$^{136}$Xe &        2.19       &  1.71 - 3.53  &  2.38 - 2.802   &      3.352      &      4.2       &       -      \\
$^{150}$Nd &          -        &      3.45     &       -         &  2.321 - 2.888  &      1.71      &  1.98 - 3.7  \\
\bottomrule
\end{tabular}
\caption{NMEs calculated in different frameworks. 
The method used to take into account short-range correlations is indicated in brackets. 
We have scaled the cited values to $r_0 = 1.2$ fm and  $g_A = 1.25$
(see Eq.~\eqref{eqn:mp}) to make them directly comparable. If ranges instead of single NME 
values are given then they arise due to intrinsic model details varied
in the respective publications. This table is graphically represented
in Fig.~\ref{fig:isonme}, 
the pseudo-SU(3) NME for $^{150}$Nd plotted there is 1.00 \cite{Hirsch:1994fw}.}
\label{tab:nme}
\end{table}

The phase space factor is through convention 
proportional to $g_A^4/R_A^2$ \cite{suhonen1998},
\begin{equation}
G^{0\nu} \propto \frac{g_A^4}{R_A^2},
\end{equation} 
with $R_A = r_0 A^{1/3}$ being the nuclear radius and $1 \lesssim g_A
\lesssim 1.25 $ the axial-vector coupling. 
The dependence on $R_A$ stems from the desire to make the NMEs
dimensionless. Therefore in the definition of the NMEs there is a
factor of $R_A$ which is compensated for by the factor $1/R_A^2$ in
$G^{0\nu}$. 
To resolve the issue of comparing matrix elements calculated using 
different values of $g_A$, some -- but not all -- authors define
\begin{equation}
\label{eqn:mp}
M^{\prime 0\nu} = \left (\frac {g_A}{1.25} \right )^2 M^{0\nu},
\end{equation}
thereby carrying the $g_A$ dependence from $G^{0\nu}$ to $M^{\prime
0\nu}$, i.e., 
\be
G^{0\nu} (M^{0\nu})^2 = G^{0\nu}_{1.25} (M^{\prime 0\nu})^2,
\ee
with $G^{0\nu}_{1.25} = G^{0\nu}(g_A=1.25)$. This means that these
NMEs share a common $G^{0\nu}$ factor -- that of $g_A=1.25$. 
Still one has to be careful when comparing NMEs from different groups,
since different authors take different values for $r_0$, usually
$r_0=1.1$ fm (e.g.~Ref.~\cite{tuebingen2009,tuebingen:150nd}) 
or $r_0=1.2$ fm (e.g.~Ref.~\cite{ibm:2009,caurier2009,jyvaskyla2009}). 
The NMEs are proportional to $r_0$ and therefore when comparing two
different matrix elements $M^{0\nu}_1$, $M^{0\nu}_2$, which have been 
calculated using $r_{0,1}$ and $r_{0,2}$, respectively, one has to
rescale $M^{0\nu}_2$ by $r_{0,1}/r_{0,2}$ or $M^{0\nu}_1$ by
$r_{0,2}/r_{0,1}$. 
Otherwise one introduces an error of $(r_{0,1}/r_{0,2})^2 \simeq 1.19$ in terms of 
half-life (see Eq.~\eqref{eqn:half-life}). 
A compilation of $g_A$, $r_0$ and $G^{0\nu}$ values used in different 
works can be found in Ref. \cite{smolnikov:hl2meff}. 

In addition, it is often overlooked that there are differences between
independent phase space factor calculations, which can be as high as
$\sim$13\% (see the Appendix of \cite{tuebingen:g0nu}). 
For instance, Ref.~\cite{tuebingen2009} uses phase space factors from 
\cite{tuebingen1999:g0nu}, while Ref.~\cite{jyvaskyla2009} uses the 
ones from \cite{suhonen1998}. There, $G^{0\nu}$ for the isotope
$^{136}$Xe is given as $49.7 \times 10^{-15}$ yrs$^{-1}$ and $43.1
\times 10^{-15}$ yrs$^{-1}$, respectively (we scaled them to
$g_A=1.25$ and $r_0=1.2$~fm to make them directly comparable). 
To perform a consistent comparison
between different NME calculations we will take the numerical values
for the phase space factors from Ref.~\cite{suhonen1998} when
calculating the necessary half-life sensitivities and take carefully 
into account all of the above mentioned difficulties\footnote{Note
that there is a misprint for the phase space factor of $^{100}$Mo in 
Ref.~\cite{suhonen1998}.}. 
Table \ref{tab:g0nu} shows the phase space factors used in our
calculations. All \nubb-isotopes with a $Q$-value above 2 MeV are
given. We have chosen the value $r_0=1.2$ fm throughout our
analysis. Also given in the table is the natural abundance of the
isotope in percent. 

The convention issues mentioned so far are of course different from the intrinsic 
uncertainty stemming from the nuclear physics itself. We will not get
into detail here, and refer to existing reviews available in the
literature \cite{tomoda91,suhonen1998}. A program to
reduce the uncertainty by independent experimental cross checks has
been launched \cite{Zuber:2005fu}, but it is unclear whether the
results will be available and conclusive for all interesting isotopes
at the time when the decisions on the experimental parameters have to
be taken.  

An important
point here are short-range correlations (SRC) since the contribution to
NMEs stems mainly from physics of internucleon distances 
$r \le (2 - 3)$ fm \cite{tue08:src}. There are different proposals 
how to treat SRC, namely via a Jastrow-like function
\cite{millerspencer:jastrow,tomoda91}, 
Unitary Correlation Operator Method (UCOM) \cite{feldmeier1998:ucom},
or Coupled Cluster Method (CCM)
\cite{muether99:ccm,muether2000:ccm,giusti1999:ccm,tuebingen2009}. 
For instance, the authors of Ref.~\cite{jyvaskyla2009} argue that UCOM should be
preferred over Jastrow while the authors of \cite{tuebingen2009} prefer CCM.
In this work we use the NME values calculated with UCOM or CCM SRC in
the NSM, QRPA, and GCM frameworks; the NME values  in the IBM
framework are calculated with Jastrow SRC. 
In the case of the 
PHFB model the authors used a statistical estimate of the
theoretical uncertainty by calculating NMEs with three different types
of SRC, four different parametrizations of the effective 
two-body interaction and taking the mean and the standard deviation. 
We used the NMEs derived in this manner and therefore no particular
SRC method can be assigned to them. 
With a chosen SRC method, some groups discuss additional sources of
error which arise, such as 
the set of single-particle states, the number of possible wave function
configurations, or other model details. These errors are given in some
publications, and we include them in our analysis. 
The NME values and ranges we have compiled and will be used in this
work are tabulated in Table \ref{tab:nme} and plotted in Fig.~\ref{fig:isonme}.
The values are scaled to $r_0 = 1.2$ fm and 
$g_A = 1.25$ so that they are directly comparable. The original NME
values can be found in column 3 of Table 8 of Ref.~\cite{caurier2009} 
(NSM), column 6 of Table III of Ref. \cite{tuebingen2009} and column 4
of Table II of Ref.~\cite{tuebingen:150nd} (QRPA, T\"ubingen group), 
column 6 of Table 1 of Ref.~\cite{jyvaskyla2009} (QRPA, Jyv\"askyl\"a
group), columns 2 and 3 of Table VI of Ref.~\cite{ibm:2009} (IBM),
column 5 of Table I of Ref.~\cite{Rodriguez:2010mn} (GCM), and column
3 of Table IV of Ref.~\cite{PhysRevC.82.064310} (PHFB). Regarding IBM,
the isotopes for which a range is given are calculated in 
Ref.~\cite{ibm:2009} with two sets of single-particle energies, one
extracted from experiment (``experimental''), the other from a
specific model (``theoretical''). Their span defines the given range. 
The IBM values without a range are unpublished ``experimental'' NMEs kindly provided by
Francesco Iachello. As only few calculations for $^{150}$Nd are
available, we also include the result from Ref.~\cite{Hirsch:1994fw},
which applied the pseudo-SU(3) Ansatz for the calculation, which is
suitable for deformed nuclei such as $^{150}$Nd. It gives by far the
lowest NME.

\subsection{Ruling out the inverted hierarchy\label{sec:IH}}

\begin{figure}[h!]
\begin{center}
\epsfig{file=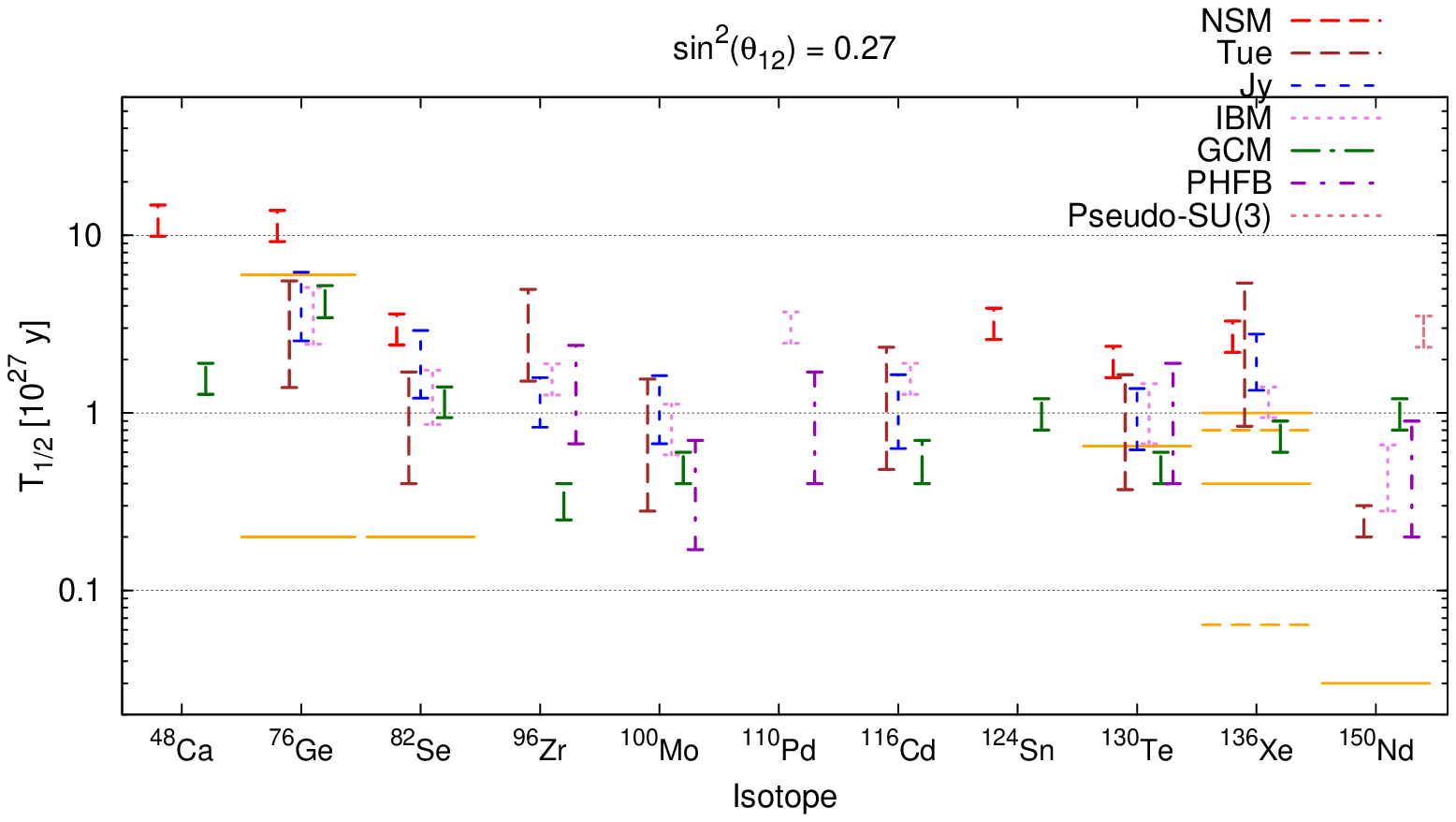,width=7.8cm,height=6.39cm} \quad
\epsfig{file=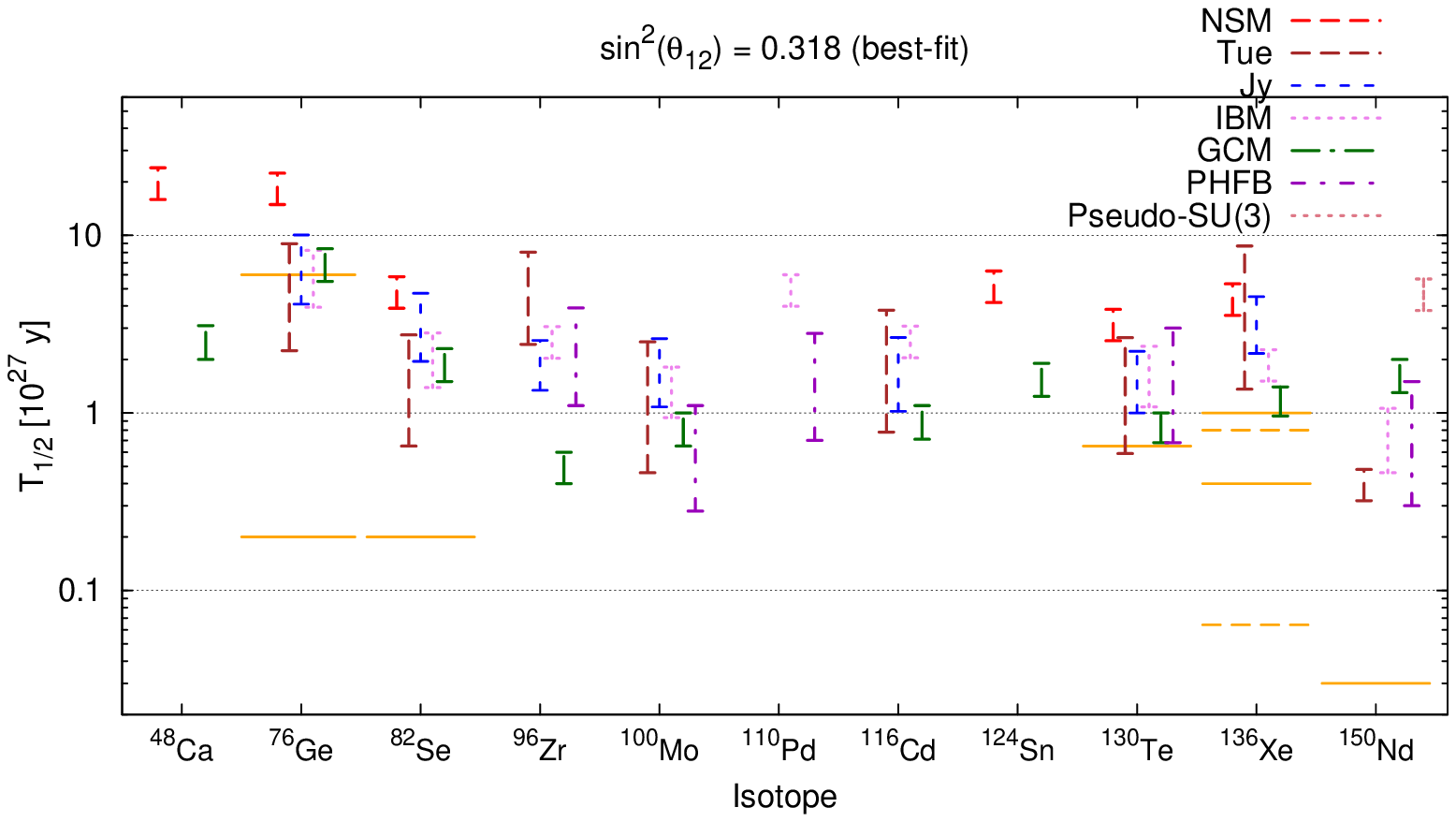,width=7.8cm,height=6.39cm}
\epsfig{file=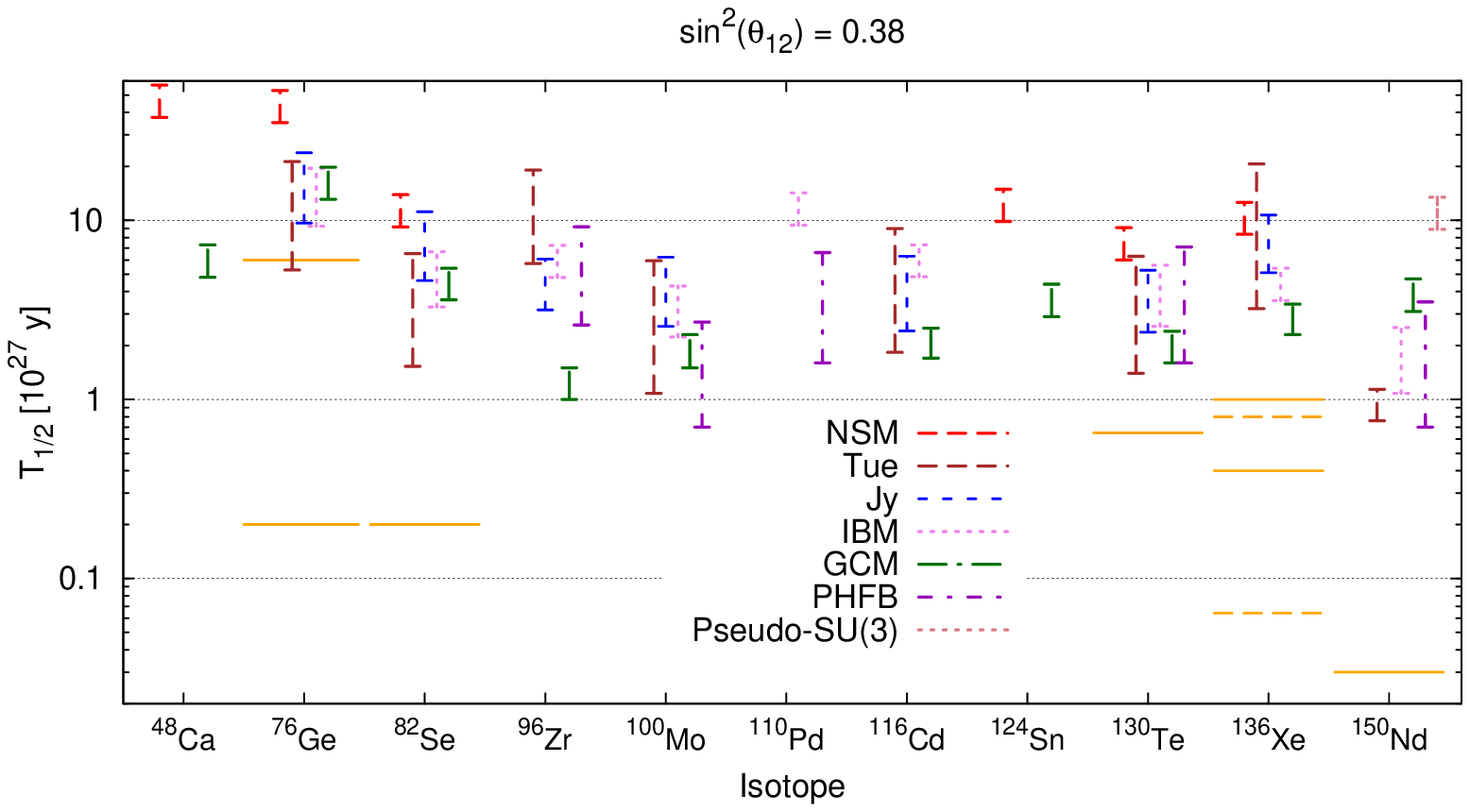,width=7.8cm,height=6.39cm} \quad
\epsfig{file=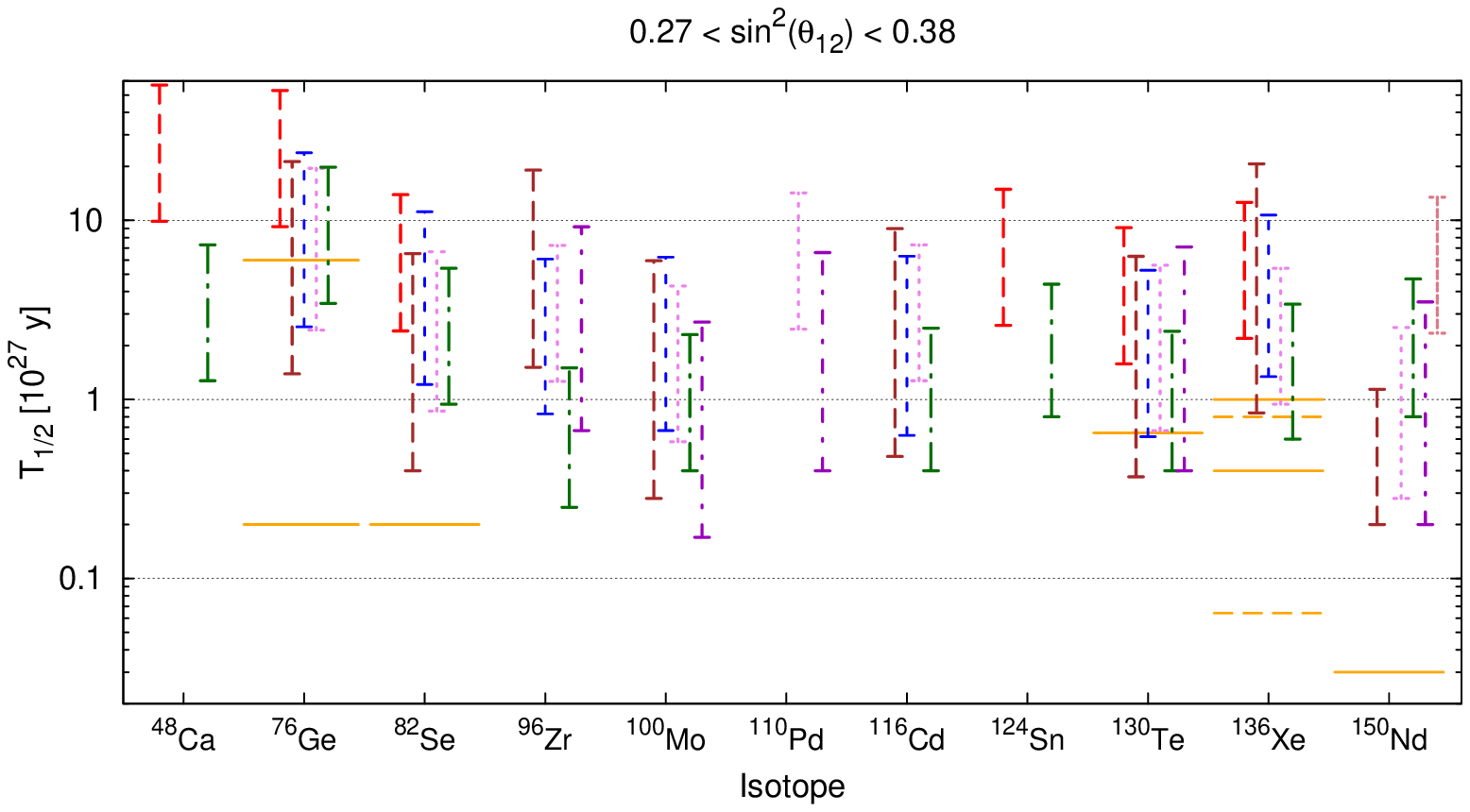,width=7.8cm,height=6.39cm}
\caption{\label{fig:isohl}Required half-life sensitivities to exclude
the inverted hierarchy for different values of $\theta_{12}$. For each value of $\sin^2 \theta_{12}$ the other
parameters ($\Delta m_{\rm A}^2$, $\Delta m_\odot^2$, $\sin^2
\theta_{13}$) are varied in their $3\sigma$ ranges.
The lower right plot tries to combine the other three: the 
lines correspond to the combined uncertainties of the nuclear physics
and the oscillation parameters. The small horizontal lines show
expected half-life sensitivities at 90\% C.L.~of running and planned
\nubb experiments. The expected limits are from the following
experiments: GERDA and MAJORANA ($^{76}$Ge, equal sensitivity
expectations for both experiments); SuperNEMO ($^{82}$Se), CUORE
($^{130}$Te); EXO ($^{136}$Xe, dashed lines); KamLAND ($^{136}$Xe,
solid lines); SNO+ ($^{150}$Nd). When two sensitivity expectations are
given for one experiment they correspond to near and far time goals.}
\end{center}
\end{figure}

\begin{table}[h!]
\centering
\begin{tabular}{@{}llc@{}}
\toprule
Isotope & Experiment & $T_{1/2}^{0\nu}$/yrs \\
\midrule
$^{76}$Ge  &  GERDA   &    $ 2.0 \times 10^{26}$\\
           &  (+MAJORANA)      &    $ 6.0 \times 10^{27}$\\
$^{82}$Se  &  SuperNEMO   &    $ 2.0 \times 10^{26}$\\
$^{130}$Te  &  CUORE  &    $ 6.5 \times 10^{26}$\\
$^{136}$Xe  &  EXO  &    $ 6.4 \times 10^{25}$\\
            &   &    $ 8.0 \times 10^{26}$\\
$^{136}$Xe  &  KamLAND  &    $ 4.0 \times 10^{26}$\\
            &   &    $ 1.0 \times 10^{27}$\\
$^{150}$Nd  &  SNO+  &    $ 4.5 \times 10^{24}$\\
            &   &    $ 3.0 \times 10^{25}$\\
\bottomrule
\end{tabular}
\caption{Expected half-life sensitivities for some \nubb experiments
\cite{Barabash:2011fn}. When two values are given they correspond to
near and far time expectations with different detector masses. }
\label{tab:fut-hl}
\end{table}

\begin{figure}[h!]
\begin{center}
\epsfig{file=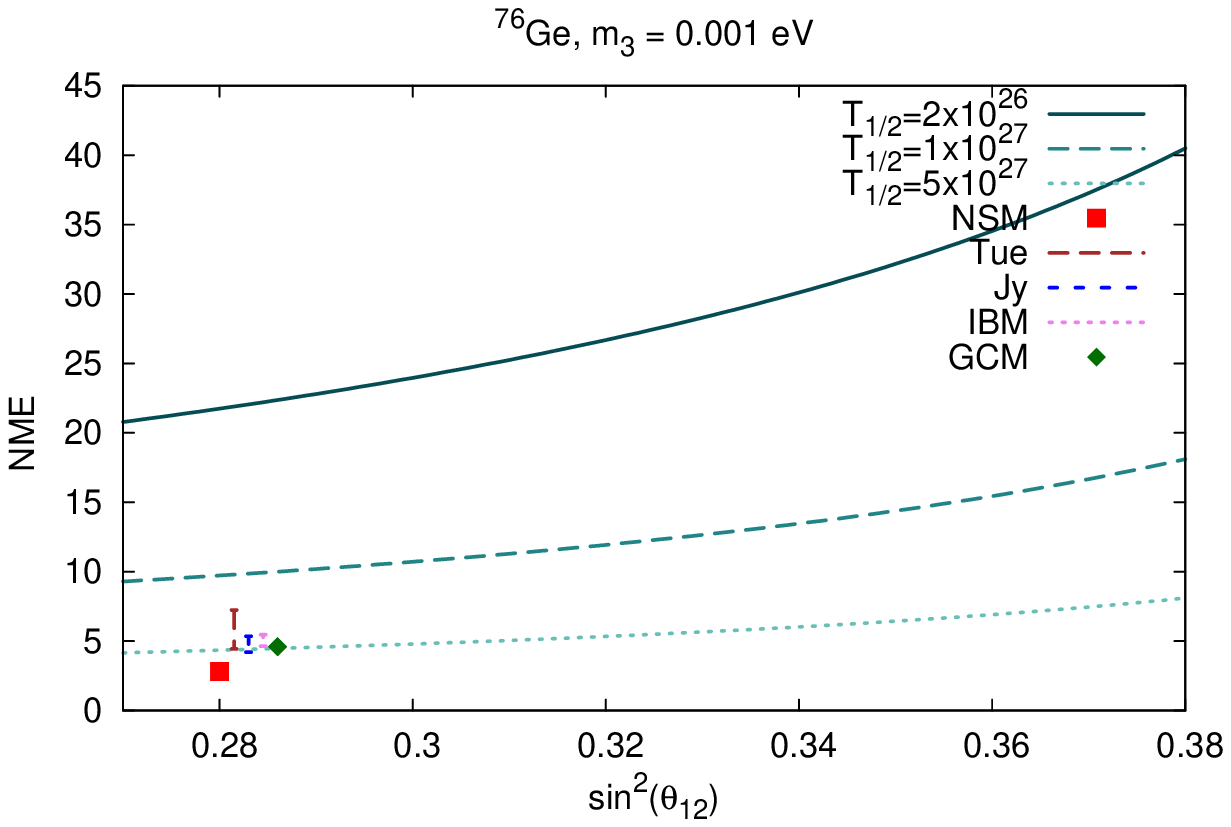,width=7.8cm,height=6.39cm} \quad
\epsfig{file=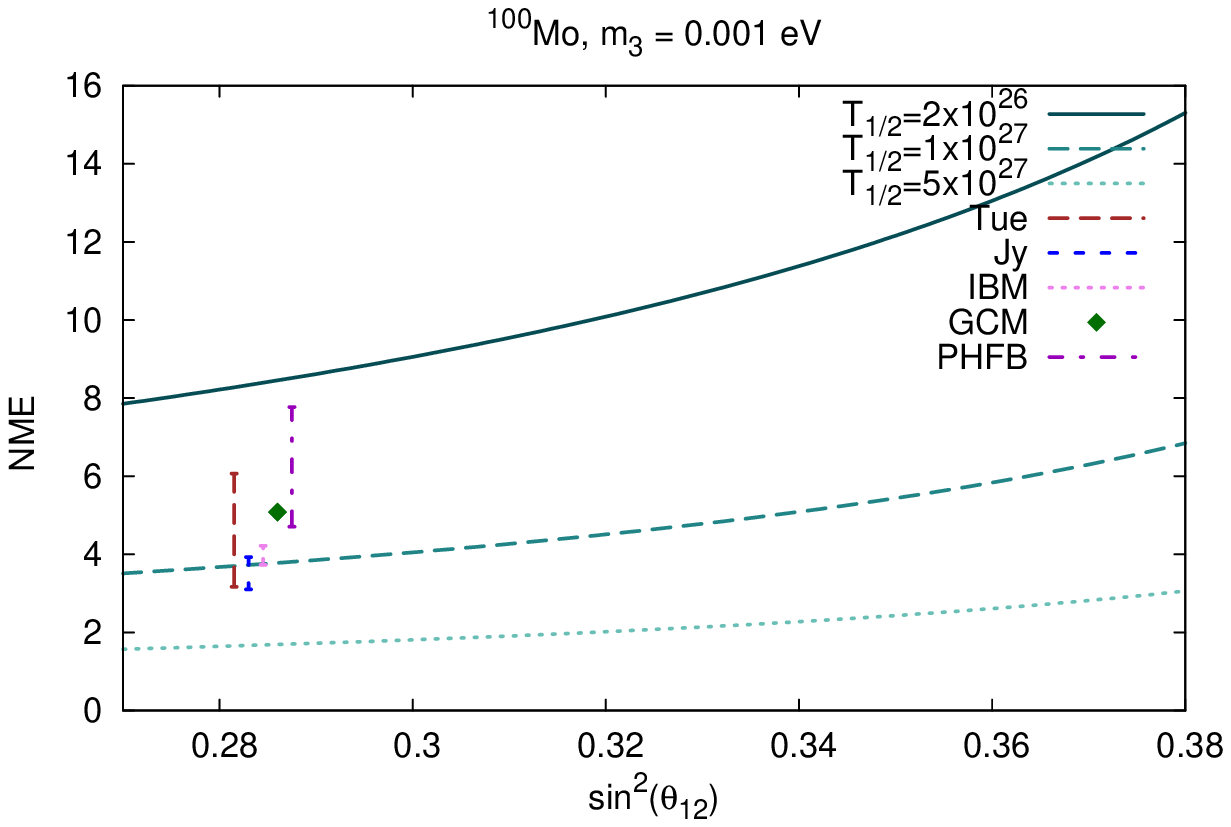,width=7.8cm,height=6.39cm}
\epsfig{file=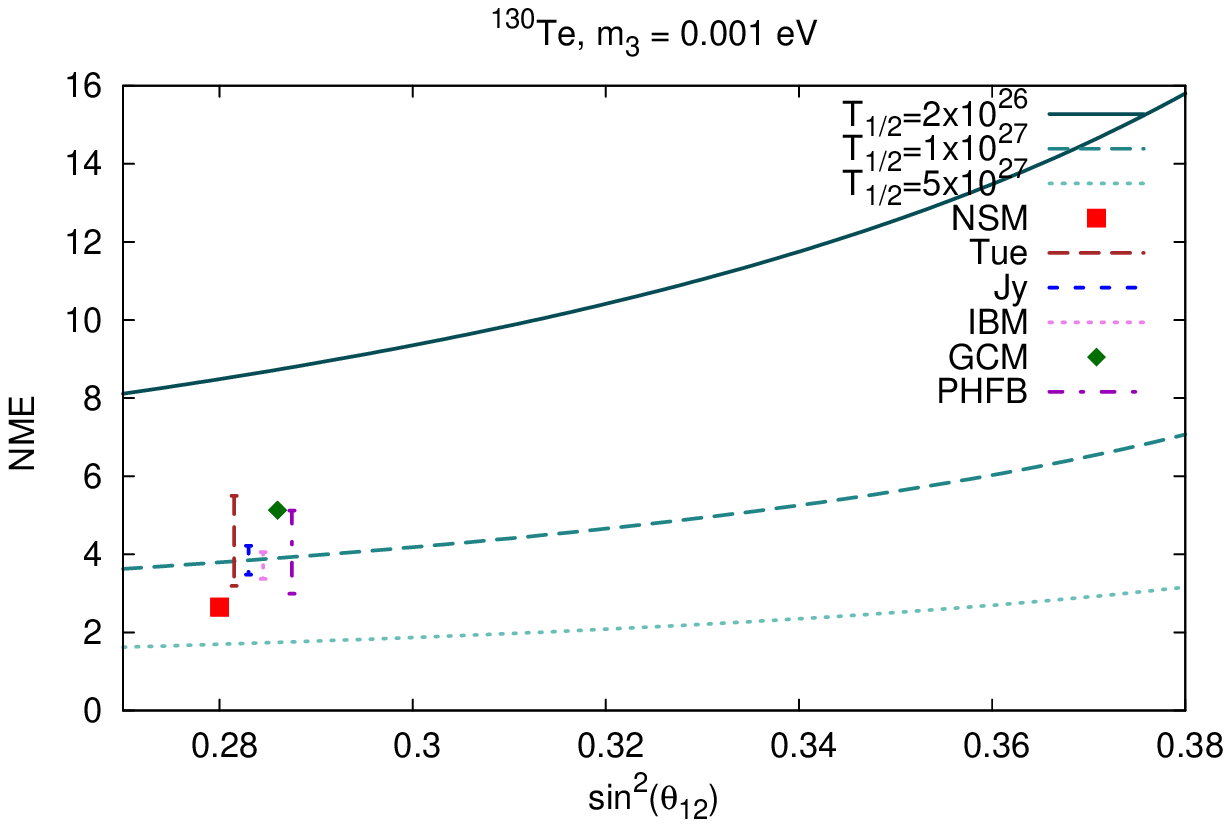,width=7.8cm,height=6.39cm} \quad
\epsfig{file=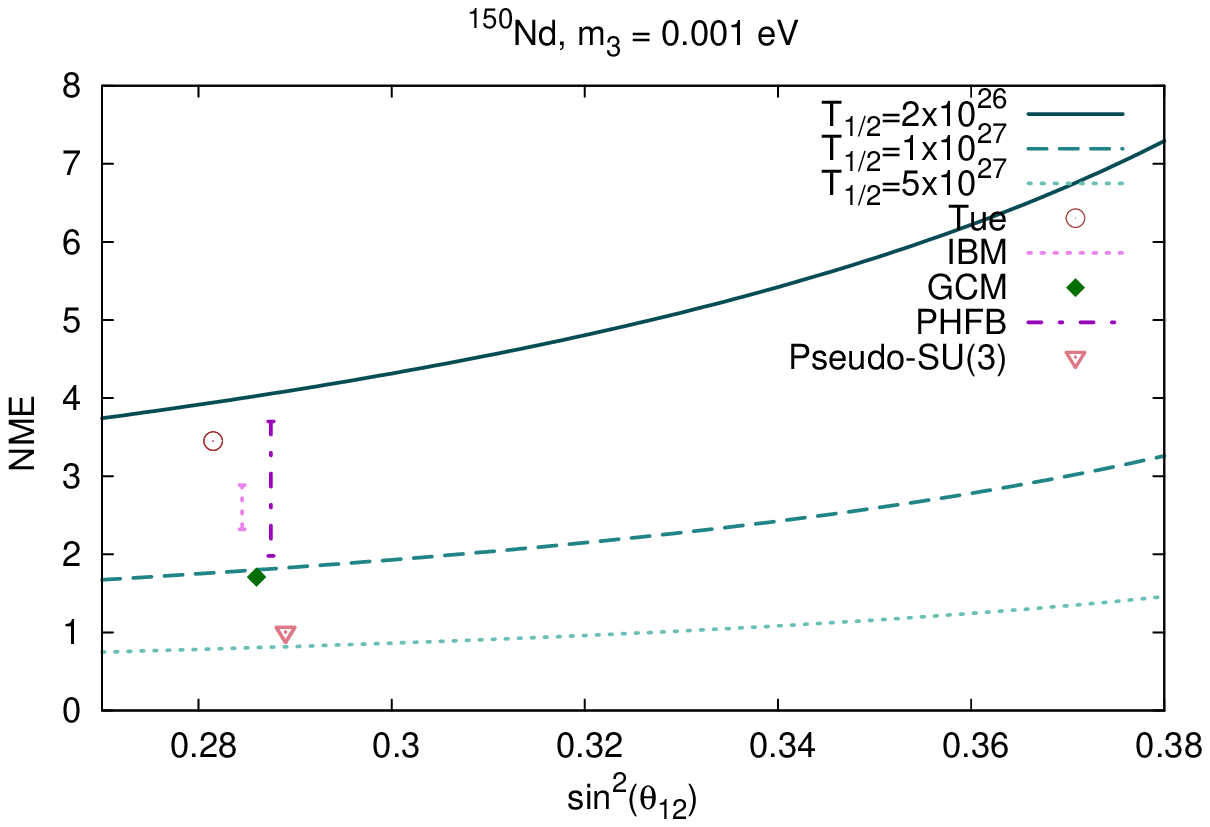,width=7.8cm,height=6.39cm}
\caption{\label{fig:nmet12}Assuming the case of an inverted neutrino
mass hierarchy and a measurement of a \nubbd signal, this plot shows
the minimal NMEs for which the inverted hierarchy can be ruled out. For the mixing parameters $\Delta m_{\rm A}^2$, $\Delta m_\odot^2$, and $\sin^2\theta_{13}$ best-fit values are taken. The ranges of
the NME calculations are also displayed in the figures.}
\end{center}
\end{figure}

\begin{figure}[h!]
\begin{center}
\epsfig{file=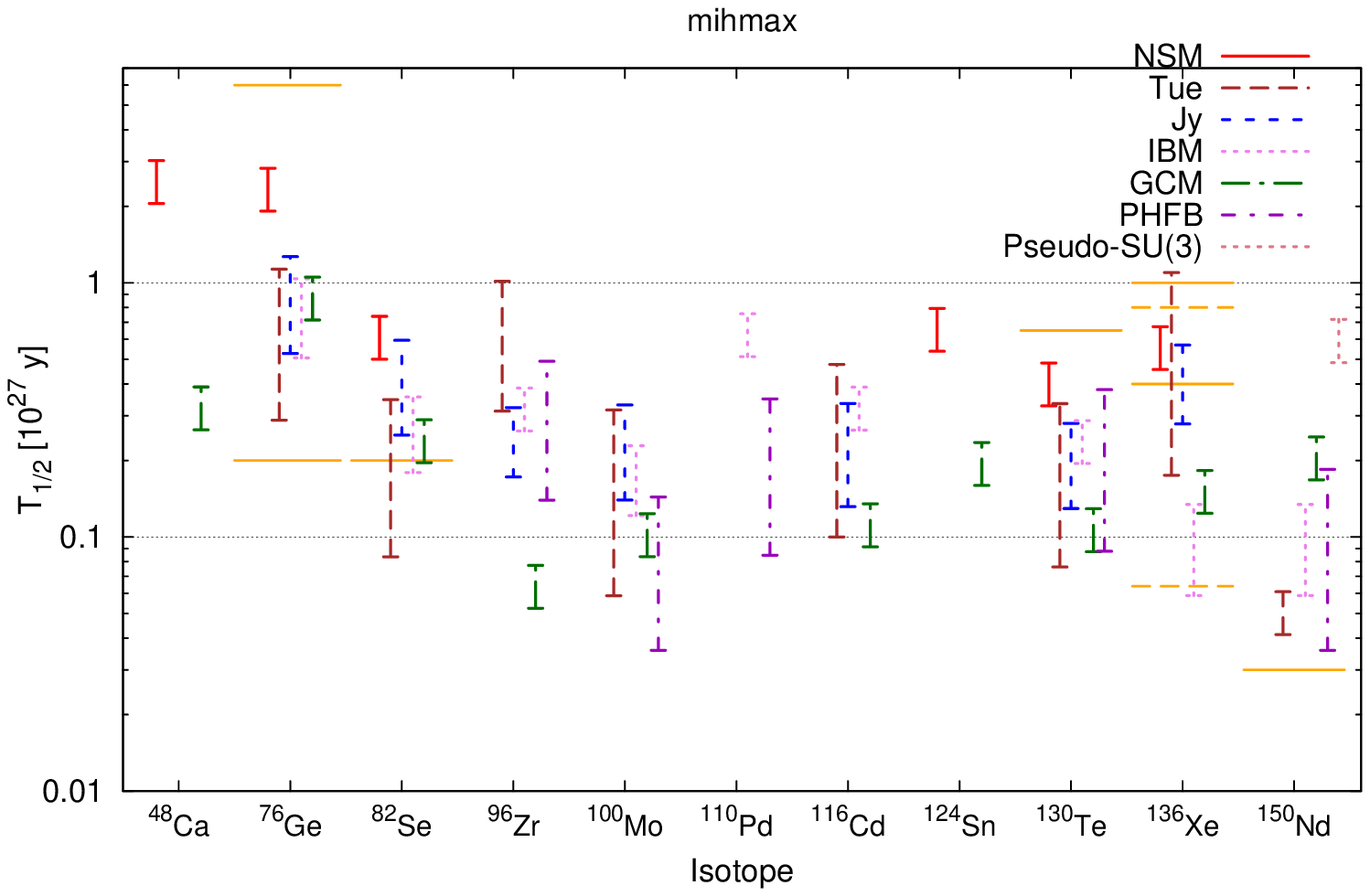,width=8.5cm,height=6.39cm}
\caption{\label{fig:isoti}Required half-life sensitivities to touch 
the inverted hierarchy. The mixing
parameters  are varied in their $3\sigma$ ranges. The small horizontal
lines show expected half-life sensitivities as in Fig. \ref{fig:isohl}.}
\end{center}
\end{figure}

Having compiled the NMEs in a form which makes it possible to compare
them with each other, we can now give the necessary half-lifes in order to rule out the
inverted hierarchy. Recall that the value 
$\langle m_\nu \rangle^{\text{IH}}_{\text{min}}$ given in
Eq.~(\ref{eqn:meffminih}) has to be reached for this, and that a
strong dependence on $\theta_{12}$ is present.  

In Fig.~\ref{fig:isohl} we plot the necessary half-lifes to rule out
the inverted hierarchy for all 11 isotopes with $Q$-value above 2
MeV. We display the situation for different values of $\theta_{12}$,
which correspond to the best-fit value of the current
oscillation analyses, and the lower and upper limit of the current
$3\sigma$ range. The full range, leaving $\theta_{12}$ free within its
current range, is also displayed. 
For convenience, we give the numerical values for necessary $T^{0\nu}_{1/2}$ 
in Table \ref{tab:half-life}, which can be
found in the Appendix. For each value of $\sin^2 \theta_{12}$ the other
parameters ($\Delta m_{\rm A}^2$, $\Delta m_\odot^2$, $\sin^2
\theta_{13}$) are varied in their $3\sigma$ ranges such that in Table
\ref{tab:half-life} one has a
somewhat more optimistic and more pessimistic prediction for the
\nubbd half-life. Recall that the dependence on the oscillation
parameters other than $\theta_{12}$ is rather weak (less than 25 \%) and will be
strongly reduced in the future.

One can compare the necessary half-lifes with the foreseen
sensitivities of up-coming experiments. We refer here to the 
compilation from Ref.~\cite{Barabash:2011fn}, which listed confirmed 
sensitivities of the currently ``most developed'' experiments. Table
\ref{tab:fut-hl} gives the numbers, staged experiments have two
values. We have included those sensitivities in our plots. 
To give an example on the interpretation of the plots, with
the final sensitivity GERDA and Majorana ($6 \times 10^{27}$ yrs) could rule out 
the inverted hierarchy if $\sin^2 \theta_{12} = 0.27$ for all NMEs
except for the NSM.

Another way to display the interplay of nuclear physics,
$\theta_{12}$ and \nubb is shown in Fig.~\ref{fig:nmet12}: assuming
for four interesting isotopes a certain half-life limit, we show 
for which NME values the inverted hierarchy is ruled out. For
instance, for $^{76}$Ge and a half-life of $5\times10^{27}$ yrs, we
can rule out the inverted hierarchy if the matrix element is larger than
about 5 if $\sin^2 \theta_{12}  = 0.32$. For a half-life of $1 \times
10^{27}$ yrs, the NME has to be larger than about 12, hence not too
realistic. Nevertheless, the ranges of
the NME calculations are also displayed in the figures.

Fig.~\ref{fig:isoti} shows the required half-life to touch the
inverted hierarchy. This half-life (corresponding to the value $\langle m_\nu \rangle^{\text{IH}}_{\text{max}}$ given in
Eq.~(\ref{eqn:meffmaxih})) does not depend on $\theta_{12}$. The other
parameters, $\dma$, $\dms$ and $\theta_{13}$ are varied in their current $3\sigma$
range. The numerical values are given in Table \ref{tab:touch-ih}. For
instance, the combined GERDA and Majorana results, as well as CUORE,
could touch the inverted hierarchy for all available NMEs.

From the figures and tables presented in this Section, one identifies
$^{100}$Mo as the somewhat most interesting isotope. With our
compilation of NMEs, the required
lifetimes to reach and/or exclude the inverted hierarchy tends to be generally
lowest for this \nubb-candidate. If the very low pseudo-SU(3) NME for
$^{150}$Nd would be omitted, then this isotope would even more
favorable than $^{100}$Mo. 
These tentative conclusions may be helpful for
experiments which have alternatives in the isotopes to investigate,
such as LUCIFER \cite{lucifer} (currently considering 
$^{82}$Se or $^{100}$Mo or $^{116}$Cd), MOON \cite{Ejiri:1999rk} 
($^{82}$Se or $^{100}$Mo), or SuperNEMO \cite{supernemo}
($^{82}$Se, $^{150}$Nd or others). 

\subsection{Current and future limits on the effective
mass\label{sec:fut}}

\begin{table}
\centering
\begin{tabular}{@{}lcccccc@{}}
\toprule
 & & & \multicolumn{2}{c}{$M^{\prime 0\nu}$} & \multicolumn{2}{c}{$\langle m_\nu \rangle$ [eV]} \\ \cmidrule(l){4-5} \cmidrule(l){6-7}
Isotope & $T_{1/2}^{0\nu}$/yrs & Experiment & min & max & min & max \\
\midrule
$^{48}$Ca  &   5.8 $\times 10^{22}$  &   CANDLES \cite{PhysRevC.78.058501}  &   0.85  &   2.37  &   3.55  &   9.91 \\
$^{76}$Ge  &   1.9 $\times 10^{25}$  &   HDM \cite{KlapdorKleingrothaus:2000sn} &   2.81  &   7.24  &   0.21  &   0.53 \\
$^{82}$Se  &   3.2 $\times 10^{23}$  &   NEMO-3 \cite{Barabash:2010bd} &   2.64  &   6.46  &   0.85  &   2.08 \\
$^{96}$Zr  &   9.2 $\times 10^{21}$  &   NEMO-3 \cite{Argyriades2010168} &   1.56  &   5.65  &   3.97  &   14.39 \\
$^{100}$Mo  &   1.0 $\times 10^{24}$  &   NEMO-3 \cite{Barabash:2010bd} &   3.10  &   7.77  &   0.31  &   0.79 \\
$^{116}$Cd  &   1.7 $\times 10^{23}$  &   SOLOTVINO \cite{PhysRevC.68.035501} &   2.51  &   4.72  &   1.22  &   2.30 \\
$^{130}$Te  &   2.8 $\times 10^{24}$  &   CUORICINO \cite{Arnaboldi2008} &   2.65  &   5.50  &   0.27  &   0.57 \\
$^{136}$Xe  &   5.0 $\times 10^{23}$  &   DAMA \cite{Bernabei200223} &   1.71  &   4.20  &   0.83  &   2.04 \\
$^{150}$Nd  &   1.8 $\times 10^{22}$  &   NEMO-3 \cite{Argyriades2008} &   1.71  &   3.70  &   2.35  &   8.65 \\ \bottomrule
\end{tabular}
\caption{Experimental \nubbd half-life limits at 90 \% C.L. 
Columns 4 and 5 show the minimal and maximal NMEs from our compilation
(see Table \ref{tab:nme}), and columns 6 and 7 the corresponding upper limits on
the effective electron neutrino mass $\langle m_\nu \rangle$. Similar
limits on $^{76}$Ge to the ones in \cite{KlapdorKleingrothaus:2000sn} 
 have been obtained by the IGEX experiment \cite{Aalseth:2002rf}.}
\label{tab:meff-lim}
\end{table}

In Table \ref{tab:meff-lim} we show the current limits on the
half-life of \nubb, obtained in a variety of experiments\footnote{Part of the
 Heidelberg-Moscow collaboration has claimed
observation \cite{Klapdor} of \nubb corresponding to a half-life of
$2.23 \times 10^{25}$ yrs, 
and a 95\% C.L.~range of $(0.8 - 18.3) \times 10^{25}$ yrs. This would
correspond to a range of the effective mass of $(0.19 - 0.49)$ eV, and 
$(0.066 - 0.82)$ eV, respectively.}. Using the
largest and smallest NME from our compilation, we give the range of
the current limit of $\meff$ for the particular isotope.

\begin{figure}[h!]
\begin{center}
\epsfig{file=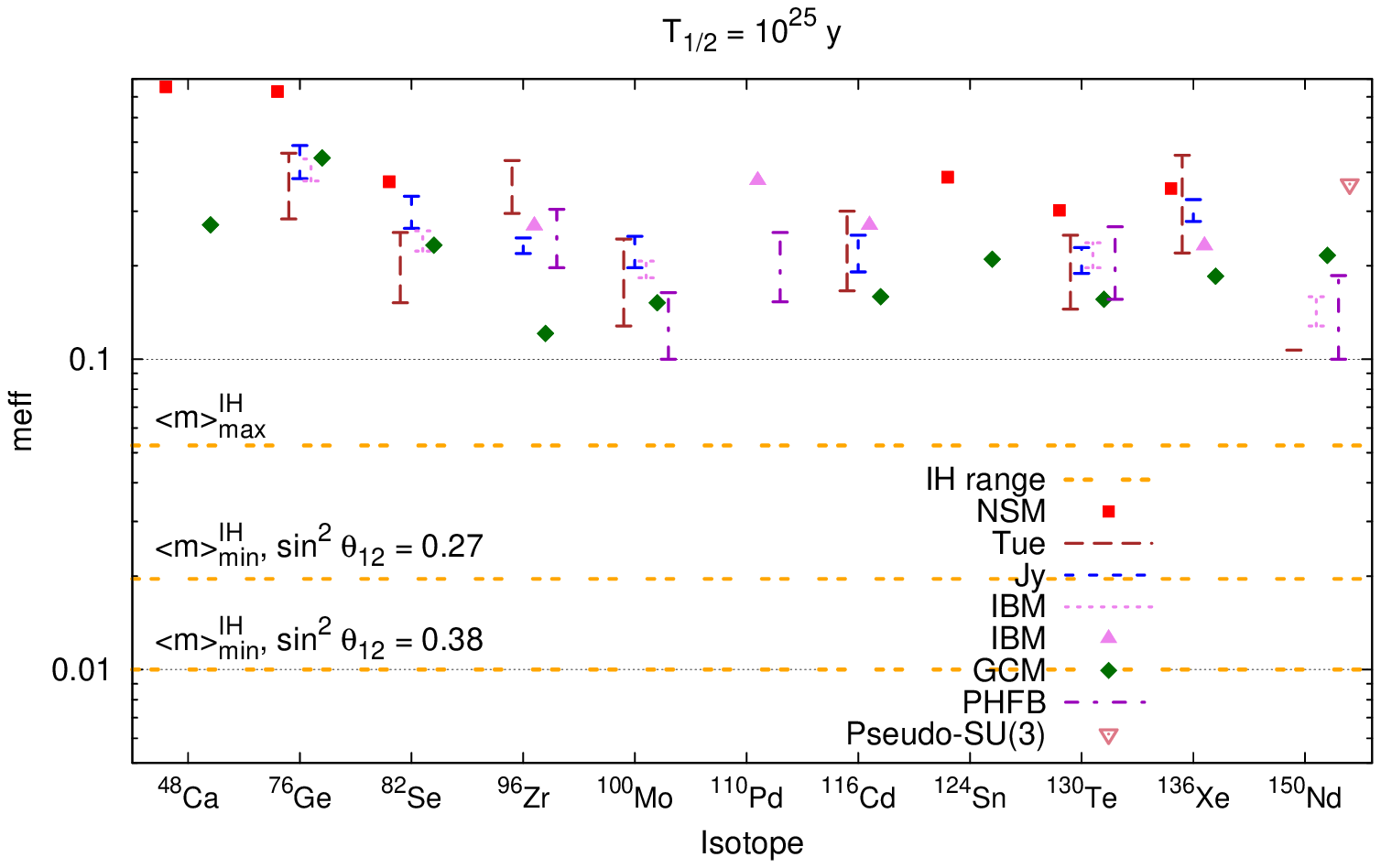,width=7.8cm,height=6.3cm} \quad
\epsfig{file=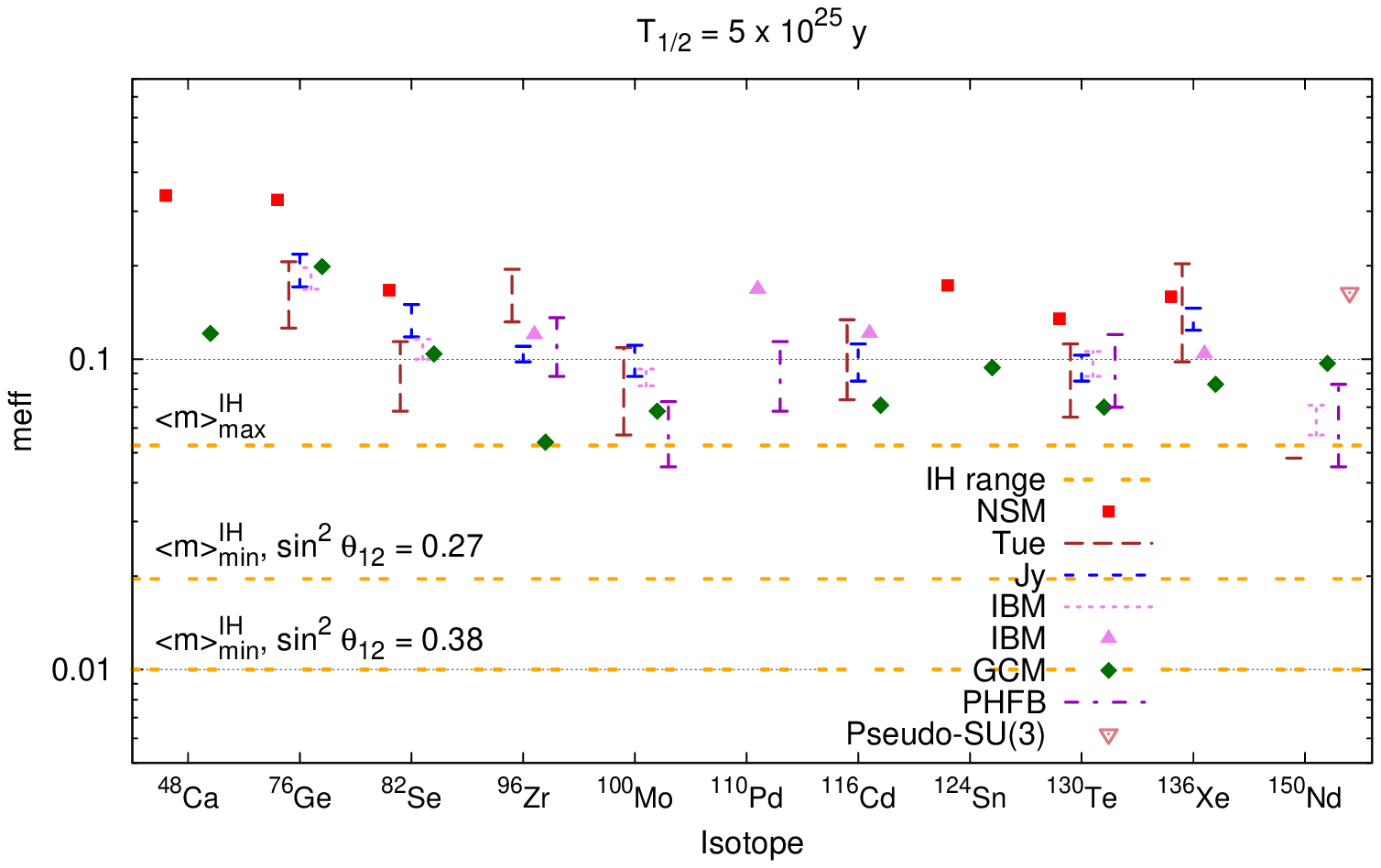,width=7.8cm,height=6.39cm}
\epsfig{file=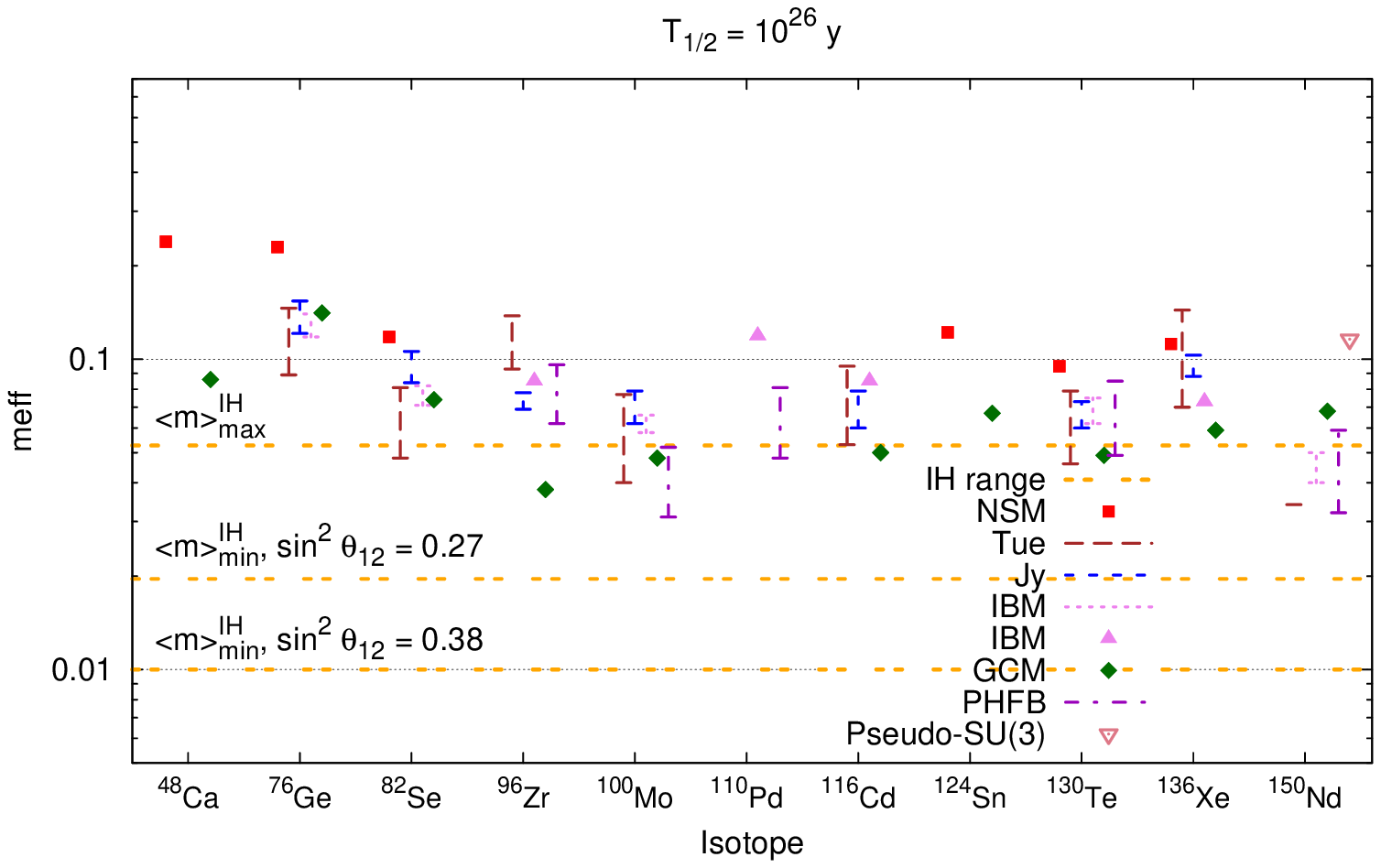,width=7.8cm,height=6.39cm} \quad
\epsfig{file=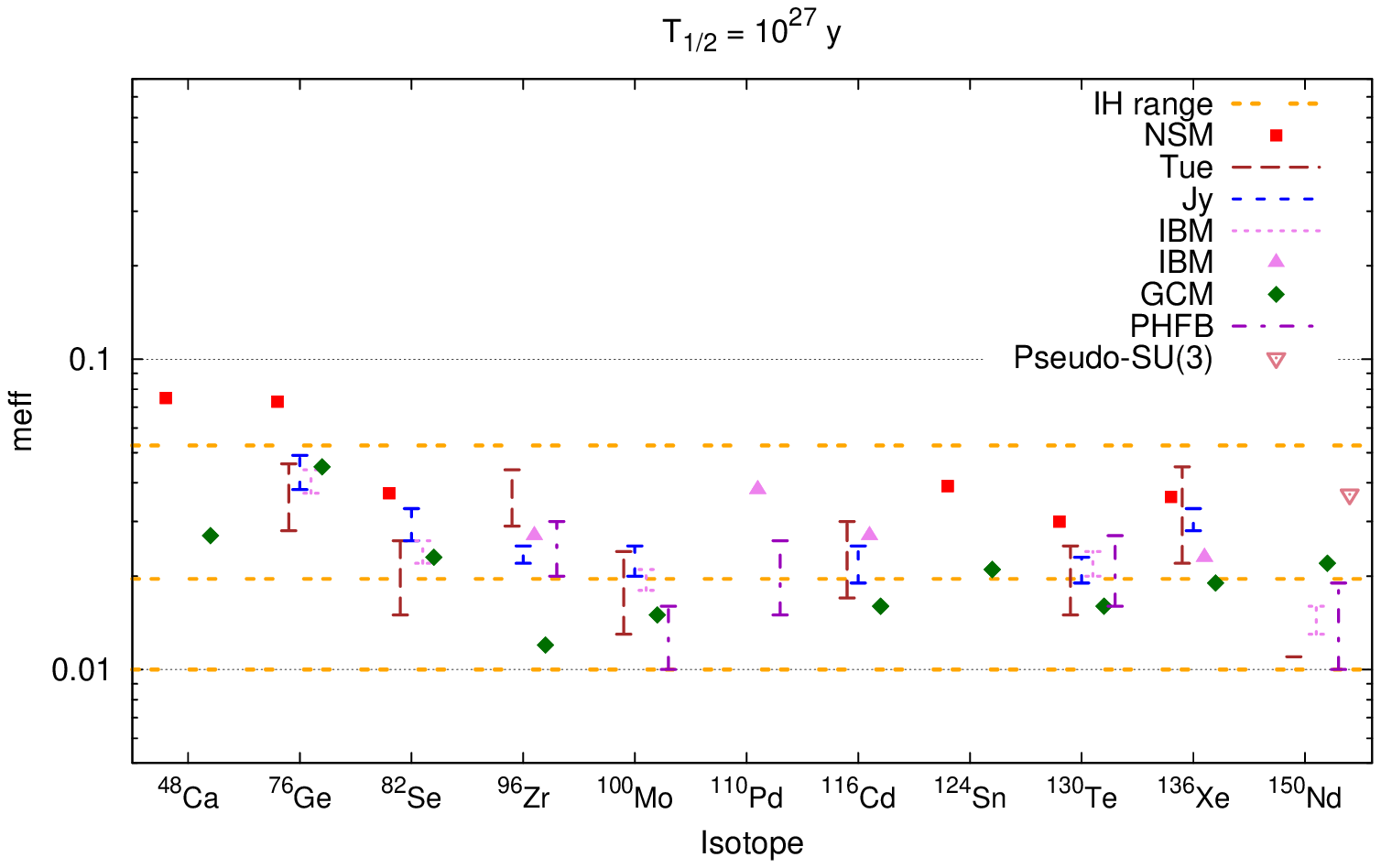,width=7.8cm,height=6.39cm}
\caption{\label{fig:gen}Limits on the effective mass which can be set
assuming different half-lifes for the 11 isotopes under
investigation. The orange double-dashed horizontal lines show the
upper and lower lines of the inverted hierarchy when the mixing parameters are
varied in their 3$\sigma$ range (see Fig.~\ref{3figs}). Thereby the
horizontal line at 0.02 eV (0.01 eV) corresponds to the lower line of
the IH for $\sin^2\theta_{12} = 0.27$ ($\sin^2\theta_{12} = 0.38$).}
\end{center}
\end{figure}

Finally, we give the limit on the effective mass as a function of 
achieved half-life for the 11 isotopes under investigation. This is
shown in Fig.~\ref{fig:gen}. We have given four different half-life 
values. 
With a
half-life sensitivities of about $5\times 10^{25}$ yrs the first isotopes
start to touch the inverted hierarchy. 
Without specifying the value of $\theta_{12}$, no
isotope can rule out the inverted hierarchy unless sensitivities above
$10^{27}$ yrs are reached. Entering the inverted hierarchy regime
requires sensitivities above $10^{26}$ yrs.

\section{Experimental consideration\label{sec:exp}}

A large variety of different upcoming experiments exists in various
stages of realization. They are in order of increasing isotope mass
CANDLES \cite{candles} ($^{48}$Ca), GERDA \cite{Abt:2004yk} and MAJORANA
\cite{Aalseth:2004yt} ($^{76}$Ge), LUCIFER \cite{lucifer} ($^{82}$Se
or $^{100}$Mo or $^{116}$Cd), SuperNEMO \cite{supernemo} ($^{82}$Se or
$^{150}$Nd), 
MOON \cite{Ejiri:1999rk} ($^{82}$Se or $^{100}$Mo), 
COBRA \cite{Zuber2001} ($^{116}$Cd) , CUORE \cite{Arnaboldi2004} ($^{130}$Te), 
EXO \cite{Danilov2000}, XMASS \cite{xmass}, KamLAND-Zen
\cite{Terashima:2008zz} and NEXT \cite{next} ($^{136}$Xe), 
DCBA \cite{dcba} and SNO+ \cite{Chen:2008un} ($^{150}$Nd). 
As discussed before not for all proposals the final decision on the selected isotope is already made.
From the discussion of the previous sections it would of course be desirable to rule out the
inverted scenario and thus tune the experimental parameters and hence
the sensitivity to do so. 
The  obtainable half-life can be estimated to be 
\be
\label{eqn:bgfree}
T_{1/2}^{0\nu}  = \frac{N_A \, \text{ln 2} }{n_\sigma} \left ( \frac{a \times \epsilon}{W} \right ) M \times t
\ee
in a background-free scenario and 
\be
\label{eqn:bglimited}
T_{1/2}^{0\nu} = \frac{N_A \, \text{ln 2} }{n_\sigma} 
\left ( \frac{a \times \epsilon}{W} \right ) \,
\sqrt{\frac{M \times t}{B \times  \Delta E}} 
\ee
 in case of a background limited search \cite{Avignone:2007fu}, where 
$n_\sigma$ is the number of standard deviations corresponding to the
desired confidence level, 
$W$ the molecular weight of the source material, 
and the other parameters as in Eq.~\eqref{eqn:hlsensbg}.
Notice, only in the background-free scenario the half-life sensitivity
scales linearly with the measuring time.
To be more conservative and realistic we assume a background limited case.
As shown in Fig.~\ref{fig:meff-t12}, the minimal effective Majorana mass which
has to be explored shows a factor of
two difference due to the current uncertainty in the mixing angle
$\theta_{12}$, depending on whether the actual value of
$\theta_{12}$ comes off at the high or low
end of its currently allowed range.  Thus, this implies a factor of 16 difference in the 
combination of measuring time, energy resolution, background index
and detector mass. 
From the experimental point of view such a big potential factor
causes a significant challenge and work,
as half-life measurements well beyond $10^{26}$ yrs itself are already
non-trivial. Therefore it would be extremely desirable to reduce the uncertainty on 
$\theta_{12}$ in future solar experiments like SNO+. \\
As an example consider a 1 ton Ge-experiment, enriched to 90 \% in $^{76}$Ge.
Furthermore, consider a full detection efficiency, an energy
resolution (FWHM) of 3 keV at peak position, the best one 
of all considered double beta experiments, and 10 years
of running time. For an optimistic combination of the other mixing
parameters (upper rows in Tables 
\ref{tab:half-life} and \ref{tab:touch-ih}) such an experiment could
touch the IH at $2\sigma$ C.L.~even using the less favorable NME if a background
level of $5.5 \times 10^{-3}$ counts/keV/kg/yr
could be achieved. This should be feasible as already for GERDA phase II
the aim is to achieve $10^{-3}$ counts/keV/kg/yr. 
Ruling out the complete IH for small $\theta_{12}$
would require $2.4 \times 10^{-4}$ counts/keV/kg/yr. For large
$\theta_{12}$ it is not possible to exclude the IH (assuming the
smallest NME) with the considered experimental parameters. Excluding
the IH would require a half-life sensitivity of $3.5 \times 10^{28}$
yrs (Tab.~\ref{tab:half-life}). But even considering a background-free
case and therefore using Eq.~\eqref{eqn:bgfree}, one obtains a
2$\sigma$ half-life limit of only $2.47 \times 10^{28}$ yrs. 
By using Eq.~\eqref{eqn:bglimited} one could formally calculate a
necessary background of $1.6 \times 10^{-5}$ counts/keV/kg/yr to
exclude a half-life of $3.5 \times 10^{28}$ yrs with the stated
experimental set-up. But this background corresponds to only about 0.5
total background counts during the whole operational period of 10
years which has to be compared to 1.4 expected counts from the
$0\nu\beta\beta$ decay with a half-life of $3.5 \times 10^{28}$ yrs. 
Hence, the experiment cannot be considered to be background dominated
and thus Eq.~\eqref{eqn:bglimited} is not applicable. 

As another example, consider a large scale experiment like SNO+ using $^{150}$Nd,
enriched to 60\%. With a total mass of 760 kg of natural Nd, 10 years of
running time and an energy resolution of about 300 keV
(the resolution  depends on the percentage of Nd-loading of the scintillator, here a 
resolution of $3.5 \%/\sqrt{E}$ was assumed) 
it would require a background of $6.1 \times 10^{-4}$ 
counts/keV/kg/yr to touch the IH at $2\sigma$ C.L. 
To exclude IH a background as small
as $2.7 \times 10^{-5}$ counts/keV/kg/yr (for
small $\theta_{12}$) or $1.9 \times 10^{-6}$
counts/keV/kg/yr (for large $\theta_{12}$) would be required (note
that due to the presence of the $2\nu\beta\beta$ mode in conjunction
with the energy resolution of only 300 keV this low background is very
hard to reach and one has to scale up the other
parameters of the experiment to fully cover the IH). 
Nevertheless,  one can check that the background levels we have
estimated here still correspond to a background dominated case. 

\section{Conclusions\label{sec:concl}}


The main focus of the present paper was put on testing the inverted
neutrino mass hierarchy with neutrinoless double beta decay
experiments. The maximal and (non-zero) minimal values of the effective
mass are natural sensitivity goals for the experimental program.  

We have stressed that the mixing parameter $\theta_{12}$,
the solar neutrino mixing angle, introduces an uncertainty of a factor
of 2 on the minimal value of the effective mass. This implies an
uncertainty of a factor of $2^2 = 4$ on the lifetime and $2^4 = 16$ on
the combination of isotope mass, background level, energy resolution
and measuring time. Given the long-standing problem of nuclear matrix
element calculations we have taken a pragmatic point of view: 
to quantify the necessary half-lifes to test
and/or rule out the inverted hierarchy we have attempted to collect as
many theoretical calculations as possible, and included their errors
if available. The nuclear matrix elements we have compiled have been
put on equal footing in what regards convention issues. 
We have used our compilation of NMEs to give the current
limits on the effective mass of different isotopes, and to give the
limits on the effective mass as a function of reached half-life. 
The isotope $^{100}$Mo tends to look interesting, in the sense that with
the same lifetime limit stronger constraints on the effective mass
than for the other isotopes can be reached, an observation potentially
interesting for upcoming experiments without a final decision on which
isotope to use. 

We finish by stressing once more that a precision determination of
the solar neutrino mixing angle $\theta_{12}$ is of crucial importance
to evaluate the physics potential of neutrinoless double beta decay
experiments. A better knowledge of this parameter is desirable, and we
hope to provide here additional motivation for further studies.

\vspace{0.3cm}
\begin{center}
{\bf Acknowledgments}
\end{center}
We thank Alexander Barabash, Jason Detwiler and Jorge Hirsch for helpful remarks. 
This work was supported by the ERC under the Starting Grant 
MANITOP and by the DFG in the Transregio 27 (A.D.~and W.R.). 

\bibliographystyle{utcaps}
\bibliography{nuphysics,bb0nu}

\providecommand{\href}[2]{#2}\begingroup\raggedright\begin{thebibliography}{10}

\bibitem{tomoda91}
T.~Tomoda, ``Double beta decay,'' {\em Rep. Prog. Phys.} {\bfseries 54} (1991)
  53--126.

\bibitem{Vergados:2002pv}
J.~Vergados, ``{The Neutrinoless double beta decay from a modern
  perspective},'' \href{http://dx.doi.org/10.1016/S0370-1573(01)00068-0}{{\em
  Phys.Rept.} {\bfseries 361} (2002) 1--56},
  \href{http://arxiv.org/abs/hep-ph/0209347}{{\ttfamily hep-ph/0209347}}.

\bibitem{Avignone:2007fu}
I.~Avignone, Frank~T., S.~R. Elliott, and J.~Engel, ``{Double Beta Decay,
  Majorana Neutrinos, and Neutrino Mass},''
  \href{http://dx.doi.org/10.1103/RevModPhys.80.481}{{\em Rev.Mod.Phys.}
  {\bfseries 80} (2008) 481--516},
  \href{http://arxiv.org/abs/0708.1033}{{\ttfamily arXiv:0708.1033 [nucl-ex]}}.

\bibitem{WR}
W.~Rodejohann, ``{Neutrinoless Double Beta Decay in Particle Physics},''
  \href{http://arxiv.org/abs/1011.4942}{{\ttfamily arXiv:1011.4942 [hep-ph]}}.

\bibitem{petcov}
S.~Pascoli and S.~Petcov, ``{The SNO solar neutrino data, neutrinoless double
  beta decay and neutrino mass spectrum},''
  \href{http://dx.doi.org/10.1016/S0370-2693(02)02510-8}{{\em Phys.Lett.}
  {\bfseries B544} (2002) 239--250},
  \href{http://arxiv.org/abs/hep-ph/0205022}{{\ttfamily hep-ph/0205022}}.

\bibitem{Petcov:2005yq}
S.~Petcov, ``{Theoretical prospects of neutrinoless double beta decay},''
  \href{http://dx.doi.org/10.1088/0031-8949/2005/T121/013}{{\em Phys.Scripta}
  {\bfseries T121} (2005) 94--101},
  \href{http://arxiv.org/abs/hep-ph/0504166}{{\ttfamily hep-ph/0504166}}.

\bibitem{schwetz2010:fit}
T.~Schwetz, M.~A. Tortola, and J.~W.~F. Valle, ``Three-flavour neutrino
  oscillation update,''
  \href{http://dx.doi.org/10.1088/1367-2630/10/11/113011}{{\em New~J.~Phys.}
  {\bfseries 10} (2008) 113011},
  \href{http://arxiv.org/abs/0808.2016v3}{{\ttfamily arXiv:0808.2016v3
  [hep-ph]}}.

\bibitem{faessler:dblbeta}
A.~Faessler and F.~Simkovic, ``Double beta decay,''
  \href{http://dx.doi.org/10.1088/0954-3899/24/12/001}{{\em J. Phys. G: Nucl.
  Part. Phys.} {\bfseries 24} (1998) 2139--2178}.

\bibitem{suhonen1998}
J.~Suhonen and O.~Civitarese, ``Weak-interaction and nuclear-structure aspects
  of nuclear double beta decay,''
  \href{http://dx.doi.org/10.1016/S0370-1573(97)00087-2}{{\em Phys. Rep.}
  {\bfseries 300} (1998) 123--214}.

\bibitem{caurier2009}
J.~Men\'{e}ndez, A.~Poves, E.~Caurier, and F.~Nowacki, ``Disassembling the
  nuclear matrix elements of the neutrinoless $\beta\beta$ decay,''
  \href{http://dx.doi.org/10.1016/j.nuclphysa.2008.12.005}{{\em Nucl. Phys. A}
  {\bfseries 818} (2009) 139--151},
  \href{http://arxiv.org/abs/0801.3760v3}{{\ttfamily arXiv:0801.3760v3
  [nucl-th]}}.

\bibitem{ibm:2009}
J.~Barea and F.~Iachello, ``Neutrinoless double-$\beta$ decay in the
  microscopic interacting boson model,''
  \href{http://dx.doi.org/10.1103/PhysRevC.79.044301}{{\em Phys. Rev. C}
  {\bfseries 79} (2009) 044301}.

\bibitem{Rodriguez:2010mn}
T.~R. Rodriguez and G.~Martinez-Pinedo, ``Energy density functional study of
  nuclear matrix elements for neutrinoless $\beta\beta$ decay,''
  \href{http://dx.doi.org/10.1103/PhysRevLett.105.252503}{{\em Phys. Rev.
  Lett.} {\bfseries 105} (2010) 252503},
  \href{http://arxiv.org/abs/1008.5260}{{\ttfamily arXiv:1008.5260 [nucl-th]}}.

\bibitem{PhysRevC.82.064310}
P.~K. Rath, R.~Chandra, K.~Chaturvedi, P.~K. Raina, and J.~G. Hirsch,
  ``Uncertainties in nuclear transition matrix elements for neutrinoless
  $\beta{}\beta{}$ decay within the projected-Hartree-Fock-Bogoliubov model,''
  \href{http://dx.doi.org/10.1103/PhysRevC.82.064310}{{\em Phys. Rev. C}
  {\bfseries 82} (2010) 064310}.

\bibitem{Zuber:2005fu}
K.~Zuber, ``{Consensus report of a workshop on matrix elements for neutrinoless
  double beta decay},'' \href{http://arxiv.org/abs/nucl-ex/0511009}{{\ttfamily
  nucl-ex/0511009}}.

\bibitem{Faessler:2008xj}
A.~Faessler, G.~Fogli, E.~Lisi, V.~Rodin, A.~Rotunno, {\em et al.}, ``{QRPA
  uncertainties and their correlations in the analysis of 0 nu beta beta
  decay},'' \href{http://dx.doi.org/10.1103/PhysRevD.79.053001}{{\em Phys.Rev.}
  {\bfseries D79} (2009) 053001},
  \href{http://arxiv.org/abs/0810.5733}{{\ttfamily arXiv:0810.5733 [hep-ph]}}.

\bibitem{SNO+}
M.~C. Chen, ``{The SNO liquid scintillator project},''
\href{http://dx.doi.org/10.1016/j.nuclphysbps.2005.03.037}{{\em Nucl. Phys.
  Proc. Suppl.} {\bfseries 145} (2005) 65--68}.

\bibitem{CLEAN}
D.~N. McKinsey and K.~J. Coakley, ``{Neutrino detection with CLEAN},''
  \href{http://dx.doi.org/10.1016/j.astropartphys.2004.10.003}{{\em Astropart.
  Phys.} {\bfseries 22} (2005) 355--368},
\href{http://arxiv.org/abs/astro-ph/0402007}{{\ttfamily astro-ph/0402007}}.

\bibitem{HERON}
Y.~Huang, R.~Lanou, H.~Maris, G.~Seidel, B.~Sethumadhavan, {\em et al.},
  ``{Potential for Precision Measurement of Solar Neutrino Luminosity by
  HERON},'' \href{http://dx.doi.org/10.1016/j.astropartphys.2008.06.003}{{\em
  Astropart.Phys.} {\bfseries 30} (2008) 1--11},
  \href{http://arxiv.org/abs/0711.4095}{{\ttfamily arXiv:0711.4095
  [astro-ph]}}.

\bibitem{LENS}
{\bfseries LENS} Collaboration, R.~Raghavan, ``{LENS, MiniLENS: Status and
  outlook},'' \href{http://dx.doi.org/10.1088/1742-6596/120/5/052014}{{\em
  J.Phys.Conf.Ser.} {\bfseries 120} (2008) 052014}.

\bibitem{Bandyopadhyay:2003du}
A.~Bandyopadhyay, S.~Choubey, and S.~Goswami, ``{Exploring the sensitivity of
  current and future experiments to Theta(odot)},''
  \href{http://dx.doi.org/10.1103/PhysRevD.67.113011}{{\em Phys. Rev.}
  {\bfseries D67} (2003) 113011},
\href{http://arxiv.org/abs/hep-ph/0302243}{{\ttfamily hep-ph/0302243}}.

\bibitem{Minakata:2004jt}
H.~Minakata, H.~Nunokawa, W.~Teves, and R.~Zukanovich~Funchal, ``{Reactor
  measurement of theta(12): Principles, accuracies and physics potentials},''
  \href{http://dx.doi.org/10.1103/PhysRevD.71.013005}{{\em Phys.Rev.}
  {\bfseries D71} (2005) 013005},
  \href{http://arxiv.org/abs/hep-ph/0407326}{{\ttfamily hep-ph/0407326}}.

\bibitem{Bandyopadhyay:2004cp}
A.~Bandyopadhyay, S.~Choubey, S.~Goswami, and S.~Petcov, ``{High precision
  measurements of theta(solar) in solar and reactor neutrino experiments},''
  \href{http://dx.doi.org/10.1103/PhysRevD.72.033013}{{\em Phys.Rev.}
  {\bfseries D72} (2005) 033013},
  \href{http://arxiv.org/abs/hep-ph/0410283}{{\ttfamily hep-ph/0410283}}.

\bibitem{Petcov:2006gy}
S.~Petcov and T.~Schwetz, ``{Precision measurement of solar neutrino
  oscillation parameters by a long-baseline reactor neutrino experiment in
  Europe},'' \href{http://dx.doi.org/10.1016/j.physletb.2006.09.063}{{\em
  Phys.Lett.} {\bfseries B642} (2006) 487--494},
  \href{http://arxiv.org/abs/hep-ph/0607155}{{\ttfamily hep-ph/0607155}}.

\bibitem{metal}
A.~B. Balantekin, ``{Significance of neutrino cross-sections for
  astrophysics},'' \href{http://dx.doi.org/10.1063/1.3274141}{{\em AIP Conf.
  Proc.} {\bfseries 1189} (2009) 11--15},
\href{http://arxiv.org/abs/0909.0226}{{\ttfamily arXiv:0909.0226 [hep-ph]}}.

\bibitem{transition}
A.~Friedland, C.~Lunardini, and C.~Pena-Garay, ``{Solar neutrinos as probes of
  neutrino matter interactions},''
  \href{http://dx.doi.org/10.1016/j.physletb.2004.05.047}{{\em Phys.Lett.}
  {\bfseries B594} (2004) 347},
  \href{http://arxiv.org/abs/hep-ph/0402266}{{\ttfamily hep-ph/0402266}}.

\bibitem{ADR}
C.~H. Albright, A.~Dueck, and W.~Rodejohann, ``{Possible Alternatives to
  Tri-bimaximal Mixing},''
  \href{http://dx.doi.org/10.1140/epjc/s10052-010-1492-2}{{\em Eur. Phys. J.}
  {\bfseries C70} (2010) 1099--1110},
\href{http://arxiv.org/abs/1004.2798}{{\ttfamily arXiv:1004.2798 [hep-ph]}}.

\bibitem{Schechter82}
J.~Schechter and J.~W.~F. Valle, ``Neutrinoless double-$\beta{}$ decay in
  SU(2)$\times$U(1) theories,''
  \href{http://dx.doi.org/10.1103/PhysRevD.25.2951}{{\em Phys. Rev. D}
  {\bfseries 25} (1982) 2951--2954}.

\bibitem{lindner:t13mnu}
M.~Lindner, A.~Merle, and W.~Rodejohann, ``{Improved limit on theta(13) and
  implications for neutrino masses in neutrino-less double beta decay and
  cosmology},'' \href{http://dx.doi.org/10.1103/PhysRevD.73.053005}{{\em Phys.
  Rev.} {\bfseries D73} (2006) 053005},
\href{http://arxiv.org/abs/hep-ph/0512143}{{\ttfamily hep-ph/0512143}}.

\bibitem{ISS}
J.~Bernabeu, M.~Blennow, P.~Coloma, A.~Donini, C.~Espinoza, {\em et al.},
  ``{EURONU WP6 2009 yearly report: Update of the physics potential of Nufact,
  superbeams and betabeams},'' \href{http://arxiv.org/abs/1005.3146}{{\ttfamily
  arXiv:1005.3146 [hep-ph]}}.

\bibitem{Audi2003337}
G.~Audi, A.~H. Wapstra, and C.~Thibault, ``The 2003 atomic mass evaluation:
  (II). Tables, graphs and references,''
  \href{http://dx.doi.org/10.1016/j.nuclphysa.2003.11.003}{{\em Nuclear Physics
  A} {\bfseries 729} (2003) 337--676}.

\bibitem{bilenky2010:0nbb}
S.~M. Bilenky, ``Neutrinoless Double Beta-Decay,''  (2010) ,
  \href{http://arxiv.org/abs/1001.1946}{{\ttfamily arXiv:1001.1946 [hep-ph]}}.

\bibitem{cowell2006:scaling}
S.~Cowell, ``Scaling factor inconsistencies in neutrinoless double beta
  decay,'' \href{http://dx.doi.org/10.1103/PhysRevC.73.028501}{{\em Phys. Rev.
  C} {\bfseries 73} (2006) 028501},
  \href{http://arxiv.org/abs/nucl-th/0512012}{{\ttfamily nucl-th/0512012}}.

\bibitem{smolnikov:hl2meff}
A.~Smolnikov and P.~Grabmayr, ``Conversion of experimental half-life to
  effective electron neutrino mass in \nubbd,''
  \href{http://dx.doi.org/10.1103/PhysRevC.81.028502}{{\em Phys. Rev. C}
  {\bfseries 81} (2010) 028502}.

\bibitem{tuebingen:g0nu}
A.~Faessler, G.~L. Fogli, E.~Lisi, V.~Rodin, A.~M. Rotunno, and F.~Simkovic,
  ``Quasiparticle random phase approximation uncertainties and their
  correlations in the analysis of $0\nu\beta\beta$ decay,''
  \href{http://dx.doi.org/10.1103/PhysRevD.79.053001}{{\em Phys. Rev. D}
  {\bfseries 79} (2009) 053001}.

\bibitem{tuebingen2009}
F.~Simkovic, A.~Faessler, H.~Muther, V.~Rodin, and M.~Stauf,
  ``0$\nu\beta\beta$-decay nuclear matrix elements with self-consistent
  short-range correlations,''
  \href{http://dx.doi.org/10.1103/PhysRevC.79.055501}{{\em Phys. Rev. C}
  {\bfseries 79} (2009) 055501},
  \href{http://arxiv.org/abs/0902.0331v1}{{\ttfamily arXiv:0902.0331v1
  [nucl-th]}}.

\bibitem{tuebingen:150nd}
D.-L. Fang, A.~Faessler, V.~Rodin, and F.~Simkovic, ``Neutrinoless
  double-$\beta$ decay of $^{150}${N}d accounting for deformation,''
  \href{http://dx.doi.org/10.1103/PhysRevC.82.051301}{{\em Phys.Rev.}
  {\bfseries C82} (2010) 051301},
  \href{http://arxiv.org/abs/1009.5579v1}{{\ttfamily arXiv:1009.5579v1
  [nucl-th]}}.

\bibitem{jyvaskyla2009}
O.~Civitarese and J.~Suhonen, ``Nuclear Matrix Elements for Double Beta Decay
  in the {QRPA} approach: a critical review,''
  \href{http://dx.doi.org/10.1088/1742-6596/173/1/012012}{{\em {Journal} of
  Physics: Conference {Series}} {\bfseries 173} (2009) 012012}.

\bibitem{Hirsch:1994fw}
J.~G. Hirsch, O.~Castanos, and P.~Hess, ``{Neutrinoless double beta decay in
  heavy deformed nuclei},''
  \href{http://dx.doi.org/10.1016/0375-9474(94)00464-X}{{\em Nucl.Phys.}
  {\bfseries A582} (1995) 124--140},
  \href{http://arxiv.org/abs/nucl-th/9407022}{{\ttfamily nucl-th/9407022}}.

\bibitem{tuebingen1999:g0nu}
F.~Simkovic, G.~Pantis, J.~D. Vergados, and A.~Faessler, ``Additional nucleon
  current contributions to neutrinoless double $\beta$ decay,''
  \href{http://dx.doi.org/10.1103/PhysRevC.60.055502}{{\em Phys. Rev. C}
  {\bfseries 60} (1999) 055502}.

\bibitem{tue08:src}
F.~Simkovic, A.~Faessler, V.~Rodin, P.~Vogel, and J.~Engel, ``Anatomy of the
  $0\nu\beta\beta$ nuclear matrix elements,''
  \href{http://dx.doi.org/10.1103/PhysRevC.77.045503}{{\em Phys. Rev. C}
  {\bfseries 77} (2008) 045503},
  \href{http://arxiv.org/abs/0710.2055}{{\ttfamily arXiv:0710.2055 [nucl-th]}}.

\bibitem{millerspencer:jastrow}
G.~A. Miller and J.~E. Spencer, ``A survey of pion charge-exchange reactions
  with nuclei,'' \href{http://dx.doi.org/10.1016/0003-4916(76)90073-7}{{\em
  Ann. Phys.} {\bfseries 100} (1976) 562--606}.

\bibitem{feldmeier1998:ucom}
H.~Feldmeier, T.~Neff, R.~Roth, and J.~Schnack, ``A unitary correlation
  operator method,''
  \href{http://dx.doi.org/10.1016/S0375-9474(97)00805-1}{{\em Nucl. Phys. A}
  {\bfseries 632} (1998) 61--95},
  \href{http://arxiv.org/abs/nucl-th/9709038}{{\ttfamily nucl-th/9709038}}.

\bibitem{muether99:ccm}
H.~M\"uther and A.~Polls, ``Correlations derived from modern nucleon-nucleon
  potentials,'' \href{http://dx.doi.org/10.1103/PhysRevC.61.014304}{{\em Phys.
  Rev. C} {\bfseries 61} (1999) 014304},
  \href{http://arxiv.org/abs/nucl-th/9908002}{{\ttfamily nucl-th/9908002}}.

\bibitem{muether2000:ccm}
H.~M\"uther and A.~Polls, ``Two-body correlations in nuclear systems,''
  \href{http://dx.doi.org/10.1016/S0146-6410(00)00105-8}{{\em Prog. Part. Nucl.
  Phys.} {\bfseries 45} (2000) 243--334},
  \href{http://arxiv.org/abs/nucl-th/0001007}{{\ttfamily nucl-th/0001007}}.

\bibitem{giusti1999:ccm}
C.~Giusti, H.~M\"uther, F.~D. Pacati, and M.~Stauf, ``Short-range and tensor
  correlations in the $^{16}$O($e,e'pn$) reaction,''
  \href{http://dx.doi.org/10.1103/PhysRevC.60.054608}{{\em Phys. Rev. C}
  {\bfseries 60} (1999) 054608},
  \href{http://arxiv.org/abs/nucl-th/9903065}{{\ttfamily nucl-th/9903065}}.

\bibitem{Barabash:2011fn}
A.~S. Barabash, ``{75 years of double beta decay: yesterday, today and
  tomorrow},'' \href{http://arxiv.org/abs/1101.4502}{{\ttfamily arXiv:1101.4502
  [nucl-ex]}}.

\bibitem{lucifer}
 {\em See e.g., talks by A.~Giuliani at BEYOND 2010 or by C.~Nones at NOW 2010}
  .

\bibitem{Ejiri:1999rk}
H.~Ejiri {\em et al.}, ``{Spectroscopy of double-beta and inverse-beta decays
  from Mo-100 for neutrinos},''
  \href{http://dx.doi.org/10.1103/PhysRevLett.85.2917}{{\em Phys. Rev. Lett.}
  {\bfseries 85} (2000) 2917--2920},
  \href{http://arxiv.org/abs/nucl-ex/9911008}{{\ttfamily nucl-ex/9911008}}.

\bibitem{supernemo}
{\bfseries NEMO} Collaboration, A.~Barabash, ``{Extrapolation of NEMO technique
  to future generation of 2beta-decay experiments},''
  \href{http://dx.doi.org/10.1023/A:1015321729742}{{\em Czech.J.Phys.}
  {\bfseries 52} (2002) 575--581}.

\bibitem{PhysRevC.78.058501}
S.~Umehara {\em et al.}, ``Neutrino-less double-$\beta{}$ decay of $^{48}Ca$
  studied by $CaF_{2}$(Eu) scintillators,''
  \href{http://dx.doi.org/10.1103/PhysRevC.78.058501}{{\em Phys. Rev. C}
  {\bfseries 78} (2008) 058501},
  \href{http://arxiv.org/abs/0810.4746}{{\ttfamily arXiv:0810.4746 [nucl-ex]}}.

\bibitem{KlapdorKleingrothaus:2000sn}
H.~V. Klapdor-Kleingrothaus {\em et al.}, ``Latest Results from the
  Heidelberg-Moscow Double Beta Decay Experiment,''
  \href{http://dx.doi.org/10.1007/s100500170022}{{\em Eur. Phys. J. A}
  {\bfseries 12} (2001) 147--154},
  \href{http://arxiv.org/abs/hep-ph/0103062}{{\ttfamily hep-ph/0103062}}.

\bibitem{Barabash:2010bd}
{\bfseries NEMO} Collaboration, A.~S. Barabash and V.~B. Brudanin,
  ``{Investigation of double beta decay with the NEMO-3 detector},''
  \href{http://arxiv.org/abs/1002.2862}{{\ttfamily arXiv:1002.2862 [nucl-ex]}}.

\bibitem{Argyriades2010168}
J.~Argyriades {\em et al.}, ``Measurement of the two neutrino double beta decay
  half-life of Zr-96 with the NEMO-3 detector,'' \href{http://dx.doi.org/DOI:
  10.1016/j.nuclphysa.2010.07.009}{{\em Nuclear Physics A} {\bfseries 847}
  (2010) 168--179}, \href{http://arxiv.org/abs/0906.2694}{{\ttfamily
  arXiv:0906.2694 [nucl-ex]}}.

\bibitem{PhysRevC.68.035501}
F.~A. Danevich {\em et al.}, ``Search for $2\beta{}$ decay of cadmium and
  tungsten isotopes: Final results of the Solotvina experiment,''
  \href{http://dx.doi.org/10.1103/PhysRevC.68.035501}{{\em Phys. Rev. C}
  {\bfseries 68} (2003) 035501}.

\bibitem{Arnaboldi2008}
C.~Arnaboldi {\em et al.}, ``Results from a search for the
  $0\nu{}\beta{}\beta{}$--decay of $^{130}Te$,''
  \href{http://dx.doi.org/10.1103/PhysRevC.78.035502}{{\em Phys. Rev. C}
  {\bfseries 78} (2008) 035502},
  \href{http://arxiv.org/abs/0802.3439}{{\ttfamily arXiv:0802.3439 [hep-ex]}}.

\bibitem{Bernabei200223}
R.~Bernabei {\em et al.}, ``Investigation of $\beta\beta$ decay modes in
  $^{134}$Xe and $^{136}$Xe,'' \href{http://dx.doi.org/DOI:
  10.1016/S0370-2693(02)02671-0}{{\em Physics Letters B} {\bfseries 546} (2002)
  23--28}.

\bibitem{Argyriades2008}
{\bfseries NEMO} Collaboration, J.~Argyriades {\em et al.}, ``Measurement of
  the double-$\beta{}$ decay half-life of $^{150}$Nd and search for
  neutrinoless decay modes with the NEMO-3 detector,''
  \href{http://dx.doi.org/10.1103/PhysRevC.80.032501}{{\em Phys. Rev. C}
  {\bfseries 80} (2009) 032501},
  \href{http://arxiv.org/abs/0810.0248}{{\ttfamily arXiv:0810.0248 [hep-ex]}}.

\bibitem{Aalseth:2002rf}
{\bfseries IGEX} Collaboration, C.~Aalseth {\em et al.}, ``{The IGEX Ge-76
  neutrinoless double beta decay experiment: Prospects for next generation
  experiments},'' \href{http://dx.doi.org/10.1103/PhysRevD.65.092007}{{\em
  Phys.Rev.} {\bfseries D65} (2002) 092007},
  \href{http://arxiv.org/abs/hep-ex/0202026}{{\ttfamily hep-ex/0202026}}.

\bibitem{Klapdor}
H.~Klapdor-Kleingrothaus and I.~Krivosheina, ``{The evidence for the
  observation of neutrinoless double beta decay: The identification of
  neutrinoless double beta events from the full spectra},''
  \href{http://dx.doi.org/10.1142/S0217732306020937}{{\em Mod.Phys.Lett.}
  {\bfseries A21} (2006) 1547--1566}.

\bibitem{candles}
S.~Umehara, T.~Kishimoto, I.~Ogawa, R.~Hazama, S.~Yoshida, {\em et al.},
  ``{CANDLES for double beta decay of Ca-48},''
  \href{http://dx.doi.org/10.1088/1742-6596/39/1/093}{{\em J.Phys.Conf.Ser.}
  {\bfseries 39} (2006) 356--358}.

\bibitem{Abt:2004yk}
I.~Abt {\em et al.}, ``{A new 76Ge double beta decay experiment at LNGS},''
  \href{http://arxiv.org/abs/hep-ex/0404039}{{\ttfamily hep-ex/0404039}}.

\bibitem{Aalseth:2004yt}
{\bfseries Majorana} Collaboration, C.~E. Aalseth {\em et al.}, ``{The Majorana
  neutrinoless double-beta decay experiment},''
  \href{http://dx.doi.org/10.1134/1.1825519}{{\em Phys. Atom. Nucl.} {\bfseries
  67} (2004) 2002--2010}, \href{http://arxiv.org/abs/hep-ex/0405008}{{\ttfamily
  hep-ex/0405008}}.

\bibitem{Zuber2001}
K.~Zuber, ``COBRA--double beta decay searches using CdTe detectors,''
  \href{http://dx.doi.org/10.1016/S0370-2693(01)01056-5}{{\em Physics Letters
  B} {\bfseries 519} (2001) 1--7},
  \href{http://arxiv.org/abs/nucl-ex/0105018}{{\ttfamily nucl-ex/0105018}}.

\bibitem{Arnaboldi2004}
C.~Arnaboldi {\em et al.}, ``CUORE: a cryogenic underground observatory for
  rare events,'' \href{http://dx.doi.org/10.1016/j.nima.2003.07.067}{{\em Nucl.
  Instrum. Meth. A} {\bfseries 518} (2004) 775--798},
  \href{http://arxiv.org/abs/nucl-ex/0212053}{{\ttfamily nucl-ex/0212053}}.

\bibitem{Danilov2000}
M.~Danilov {\em et al.}, ``Detection of very small neutrino masses in
  double--beta decay using laser tagging,''
  \href{http://dx.doi.org/10.1016/S0370-2693(00)00404-4}{{\em Phys. Lett. B}
  {\bfseries 480} (2000) 12--18},
  \href{http://arxiv.org/abs/hep-ex/0002003}{{\ttfamily hep-ex/0002003}}.

\bibitem{xmass}
Y.~Takeuchi, ``{Recent status of the XMASS project},'' {\em Prepared for ICHEP
  04, Beijing, China, 16-22 Aug 2004,} (2004) 324--327.

\bibitem{Terashima:2008zz}
{\bfseries KamLAND} Collaboration, A.~Terashima {\em et al.}, ``{R $\&$ D for
  possible future improvements of KamLAND},''
  \href{http://dx.doi.org/10.1088/1742-6596/120/5/052029}{{\em
  J.Phys.Conf.Ser.} {\bfseries 120} (2008) 052029}.

\bibitem{next}
{\bfseries NEXT} Collaboration, .~F. Granena {\em et al.}, ``{NEXT, a HPGXe TPC
  for neutrinoless double beta decay searches},''
  \href{http://arxiv.org/abs/0907.4054}{{\ttfamily arXiv:0907.4054 [hep-ex]}}.

\bibitem{dcba}
N.~Ishihara, T.~Inagaki, T.~Ohama, K.~Omata, S.~Takeda, {\em et al.}, ``{A
  Separation method of 0 neutrino and 2 neutrino events in double beta decay
  experiments with DCBA},''
  \href{http://dx.doi.org/10.1016/S0168-9002(99)01003-7}{{\em
  Nucl.Instrum.Meth.} {\bfseries A443} (2000) 101--107}.

\bibitem{Chen:2008un}
{\bfseries SNO+} Collaboration, M.~C. Chen, ``{The SNO+ Experiment},''
  \href{http://arxiv.org/abs/0810.3694}{{\ttfamily arXiv:0810.3694 [hep-ex]}}.

\end{thebibliography}\endgroup


\section*{Appendix}
\begin{longtable}{@{}l@{~\,}lc@{~}c@{}c@{~}c@{}c@{~}c@{}}
\caption{Required \nubbd half-life sensitivity (in $10^{27}$ yrs)  
in order to {\it
exclude} the inverted hierarchy for different values of 
$\sin^2 \theta_{12}$. For each value of $\sin^2 \theta_{12}$ we
present two values/ranges by varying the other parameters which
determine the effective mass ($\Delta m_{\rm A}^2$, $\Delta
m_\odot^2$, $\sin^2 \theta_{13}$) in their currently allowed 3$\sigma$
region. Thereby, for each value of $\sin^2 \theta_{12}$, the numbers
in the first row correspond to the smallest possible half-lifes while
the numbers in the second row correspond to the largest half-lifes. The values calculated using the pseudo-SU(3) NME for $^{150}$Nd from Ref. \cite{Hirsch:1994fw} are (in the just described order) 2.34, 3.51, 3.77, 5.68, 8.90, 13.5 ($\times 10^{27}$ yrs).
\label{tab:half-life}}\\
\toprule
& &\multicolumn{6}{c}{half-life sensitivity to exclude IH [10$^{27}$ yrs]} \\ \cmidrule(l){3-8}
Isotope & sin$^2\theta_{12}$ & NSM \cite{caurier2009} & T\"u \cite{tuebingen2009,tuebingen:150nd} & Jy \cite{jyvaskyla2009} & IBM \cite{ibm:2009} & GCM \cite{Rodriguez:2010mn} & PHFB \cite{PhysRevC.82.064310}\\ \midrule
\endfirsthead
\caption[]{(continued)}\\
Isotope & sin$^2\theta_{12}$ & NSM & T\"ubingen & Jyv\"askyl\"a & IBM & GCM & PHFB\\ \midrule
\endhead
$^{48}$Ca  & 0.270 &      9.88       &       -       &       -       &       -         &      1.27      &       -        \\*
           &       &      14.82       &       -       &       -       &       -         &      1.91      &       -        \\*
           & 0.318 &      15.93       &       -       &       -       &       -         &      2.05      &       -        \\*
           &       &      24.00       &       -       &       -       &       -         &      3.09      &       -        \\*
           & 0.380 &      37.62       &       -       &       -       &       -         &      4.84      &       -        \\*
           &       &      56.92       &       -       &       -       &       -         &      7.32      &       -        \\* \midrule
$^{76}$Ge  & 0.270 &      9.22       &  1.39 - 3.69  &  2.54 - 4.13  &  2.44 - 3.39  &      3.44      &       -        \\*
           &       &      13.82       &  2.08 - 5.53  &  3.80 - 6.20  &  3.65 - 5.08  &      5.16      &       -        \\*
           & 0.318 &      14.86       &  2.24 - 5.95  &  4.09 - 6.67  &  3.93 - 5.46  &      5.55      &       -        \\*
           &       &      22.38       &  3.37 - 8.96  &  6.16 - 10.04  &  5.92 - 8.22  &      8.35      &       -        \\*
           & 0.380 &      35.09       &  5.28 - 14.05  &  9.66 - 15.74  &  9.28 - 12.89  &      13.09      &       -        \\*
           &       &      53.08       &  7.99 - 21.26  &  14.62 - 23.82  &  14.03 - 19.50  &      19.81      &       -        \\* \midrule
$^{82}$Se  & 0.270 &      2.41       &  0.40 - 1.13  &  1.21 - 1.94  &  0.86 - 1.16  &      0.94      &       -        \\*
           &       &      3.61       &  0.60 - 1.70  &  1.82 - 2.91  &  1.29 - 1.74  &      1.41      &       -        \\*
           & 0.318 &      3.88       &  0.65 - 1.83  &  1.95 - 3.13  &  1.39 - 1.87  &      1.52      &       -        \\*
           &       &      5.85       &  0.98 - 2.75  &  2.94 - 4.71  &  2.09 - 2.82  &      2.29      &       -        \\*
           & 0.380 &      9.17       &  1.53 - 4.31  &  4.61 - 7.39  &  3.28 - 4.42  &      3.59      &       -        \\*
           &       &      13.88       &  2.32 - 6.52  &  6.98 - 11.17  &  4.97 - 6.68  &      5.43      &       -        \\* \midrule
$^{96}$Zr  & 0.270 &        -      &  1.51 - 3.31  &  0.83 - 1.05  &      1.26      &      0.25      &  0.67 - 1.60 \\*
           &       &        -      &  2.26 - 4.96  &  1.24 - 1.58  &      1.89      &      0.38      &  1.01 - 2.41 \\*
           & 0.318 &        -      &  2.43 - 5.34  &  1.34 - 1.70  &      2.03      &      0.41      &  1.08 - 2.59 \\*
           &       &        -      &  3.66 - 8.04  &  2.01 - 2.56  &      3.06      &      0.61      &  1.63 - 3.90 \\*
           & 0.380 &        -      &  5.73 - 12.60  &  3.16 - 4.01  &      4.79      &      0.96      &  2.56 - 6.11 \\*
           &       &        -      &  8.67 - 19.06  &  4.77 - 6.07  &      7.25      &      1.45      &  3.87 - 9.24 \\* \midrule
$^{100}$Mo & 0.270 &        -      &  0.28 - 1.03  &  0.67 - 1.08  &  0.58 - 0.75  &      0.40      &  0.17 - 0.47 \\*
           &       &        -      &  0.42 - 1.55  &  1.01 - 1.62  &  0.88 - 1.12  &      0.60      &  0.26 - 0.70 \\*
           & 0.318 &        -      &  0.46 - 1.66  &  1.08 - 1.74  &  0.94 - 1.20  &      0.65      &  0.28 - 0.76 \\*
           &       &        -      &  0.69 - 2.51  &  1.63 - 2.62  &  1.42 - 1.81  &      0.98      &  0.42 - 1.14 \\*
           & 0.380 &        -      &  1.08 - 3.93  &  2.56 - 4.11  &  2.23 - 2.84  &      1.53      &  0.66 - 1.78 \\*
           &       &        -      &  1.63 - 5.94  &  3.88 - 6.22  &  3.37 - 4.30  &      2.32      &  0.99 - 2.70 \\* \midrule
$^{110}$Pd & 0.270 &        -      &       -       &       -       &      2.47      &       -        &  0.41 - 1.14 \\*
           &       &        -      &       -       &       -       &      3.70      &       -        &  0.61 - 1.71 \\*
           & 0.318 &        -      &       -       &       -       &      3.98      &       -        &  0.66 - 1.84 \\*
           &       &        -      &       -       &       -       &      5.99      &       -        &  0.99 - 2.77 \\*
           & 0.380 &        -      &       -       &       -       &      9.39      &       -        &  1.55 - 4.34 \\*
           &       &        -      &       -       &       -       &      14.21      &       -        &  2.35 - 6.57 \\* \midrule
$^{116}$Cd & 0.270 &        -      &  0.48 - 1.56  &  0.63 - 1.09  &      1.27      &      0.44      &       -        \\*
           &       &        -      &  0.72 - 2.34  &  0.95 - 1.64  &      1.90      &      0.66      &       -        \\*
           & 0.318 &        -      &  0.78 - 2.51  &  1.02 - 1.76  &      2.04      &      0.71      &       -        \\*
           &       &        -      &  1.17 - 3.79  &  1.54 - 2.66  &      3.08      &      1.07      &       -        \\*
           & 0.380 &        -      &  1.83 - 5.93  &  2.41 - 4.16  &      4.83      &      1.68      &       -        \\*
           &       &        -      &  2.77 - 8.98  &  3.65 - 6.30  &      7.30      &      2.54      &       -        \\* \midrule
$^{124}$Sn & 0.270 &      2.59       &       -       &       -       &       -         &      0.77      &       -        \\*
           &       &      3.88       &       -       &       -       &       -         &      1.15      &       -        \\*
           & 0.318 &      4.18       &       -       &       -       &       -         &      1.24      &       -        \\*
           &       &      6.29       &       -       &       -       &       -         &      1.87      &       -        \\*
           & 0.380 &      9.86       &       -       &       -       &       -         &      2.93      &       -        \\*
           &       &      14.92       &       -       &       -       &       -         &      4.43      &       -        \\* \midrule
$^{130}$Te & 0.270 &      1.58       &  0.37 - 1.09  &  0.62 - 0.91  &  0.67 - 0.97  &      0.42      &  0.42 - 1.24 \\*
           &       &      2.37       &  0.55 - 1.64  &  0.93 - 1.37  &  1.01 - 1.46  &      0.63      &  0.63 - 1.86 \\*
           & 0.318 &      2.55       &  0.59 - 1.76  &  1.00 - 1.47  &  1.08 - 1.57  &      0.68      &  0.68 - 2.00 \\*
           &       &      3.83       &  0.89 - 2.65  &  1.51 - 2.22  &  1.63 - 2.37  &      1.02      &  1.03 - 3.01 \\*
           & 0.380 &      6.01       &  1.40 - 4.16  &  2.37 - 3.48  &  2.56 - 3.71  &      1.60      &  1.61 - 4.72 \\*
           &       &      9.09       &  2.11 - 6.29  &  3.58 - 5.26  &  3.88 - 5.61  &      2.43      &  2.44 - 7.14 \\* \midrule
$^{136}$Xe & 0.270 &      2.19       &  0.84 - 3.59  &  1.34 - 1.86  &      0.94      &      0.60      &       -        \\*
           &       &      3.29       &  1.26 - 5.38  &  2.01 - 2.78  &      1.40      &      0.89      &       -        \\*
           & 0.318 &      3.54       &  1.36 - 5.78  &  2.16 - 2.99  &      1.51      &      0.96      &       -        \\*
           &       &      5.33       &  2.04 - 8.71  &  3.25 - 4.51  &      2.27      &      1.45      &       -        \\*
           & 0.380 &      8.35       &  3.21 - 13.65  &  5.10 - 7.07  &      3.56      &      2.27      &       -        \\*
           &       &      12.63       &  4.85 - 20.65  &  7.72 - 10.70  &      5.39      &      3.43      &       -        \\* \midrule
$^{150}$Nd & 0.270 &        -      &      0.20      &       -       &  0.28 - 0.44  &      0.81      &  0.17 - 0.60 \\*
           &       &        -      &      0.30      &       -       &  0.42 - 0.66  &      1.21      &  0.26 - 0.90 \\*
           & 0.318 &        -      &      0.32      &       -       &  0.46 - 0.71  &      1.30      &  0.28 - 0.97 \\*
           &       &        -      &      0.48      &       -       &  0.69 - 1.06  &      1.96      &  0.42 - 1.46 \\*
           & 0.380 &        -      &      0.76      &       -       &  1.08 - 1.67  &      3.07      &  0.66 - 2.29 \\*
           &       &        -      &      1.14      &       -       &  1.63 - 2.52  &      4.65      &  0.99 - 3.47 \\* \bottomrule
\end{longtable}


\begin{longtable}{@{}lcccccc@{}}
\caption{Same as Table \ref{tab:half-life}, but here the 
required \nubbd half-life sensitivity in order to {\it touch} the
inverted hierarchy is given. For the pseudo-SU(3) NME for $^{150}$Nd we get the values 0.49, 0.72 ($\times 10^{27}$ yrs).
\label{tab:touch-ih}}\\
\toprule
& \multicolumn{6}{c}{half-life sensitivity to touch IH [10$^{27}$ yrs]} \\ \cmidrule(l){2-7}
Isotope  & NSM \cite{caurier2009} & T\"ubingen \cite{tuebingen2009,tuebingen:150nd} & Jyv\"askyl\"a \cite{jyvaskyla2009} & IBM \cite{ibm:2009} & GCM \cite{Rodriguez:2010mn} & PHFB \cite{PhysRevC.82.064310}\\ \midrule
\endfirsthead
\caption[]{(continued)}\\
Isotope  & NSM & T\"ubingen & Jyv\"askyl\"a & IBM & GCM & PHFB\\ \midrule
\endhead
$^{48}$Ca    &        2.05       &         -       &         -       &         -         &        0.26      &         -        \\*
             &        3.03       &         -       &         -       &         -         &        0.39      &         -        \\* \midrule
$^{76}$Ge    &        1.91       &    0.29 - 0.77  &    0.53 - 0.86  &    0.51 - 0.70  &        0.71      &         -        \\*
             &        2.82       &    0.42 - 1.13  &    0.78 - 1.27  &    0.75 - 1.04  &        1.05      &         -        \\* \midrule
$^{82}$Se    &        0.50       &    0.08 - 0.24  &    0.25 - 0.40  &    0.18 - 0.24  &        0.20      &         -        \\*
             &        0.74       &    0.12 - 0.35  &    0.37 - 0.59  &    0.26 - 0.36  &        0.29      &         -        \\* \midrule
$^{96}$Zr    &          -      &    0.31 - 0.69  &    0.17 - 0.22  &        0.26      &        0.05      &    0.14 - 0.33 \\*
             &          -      &    0.46 - 1.01  &    0.25 - 0.32  &        0.39      &        0.08      &    0.21 - 0.49 \\* \midrule
$^{100}$Mo   &          -      &    0.06 - 0.21  &    0.14 - 0.22  &    0.12 - 0.16  &        0.08      &    0.04 - 0.10 \\*
             &          -      &    0.09 - 0.32  &    0.21 - 0.33  &    0.18 - 0.23  &        0.12      &    0.05 - 0.14 \\* \midrule
$^{110}$Pd   &          -      &         -       &         -       &        0.51      &         -        &    0.08 - 0.24 \\*
             &          -      &         -       &         -       &        0.76      &         -        &    0.12 - 0.35 \\* \midrule
$^{116}$Cd   &          -      &    0.10 - 0.32  &    0.13 - 0.23  &        0.26      &        0.09      &         -        \\*
             &          -      &    0.15 - 0.48  &    0.19 - 0.33  &        0.39      &        0.13      &         -        \\* \midrule
$^{124}$Sn   &        0.54       &         -       &         -       &         -         &        0.16      &         -        \\*
             &        0.79       &         -       &         -       &         -         &        0.24      &         -        \\* \midrule
$^{130}$Te   &        0.33       &    0.08 - 0.23  &    0.13 - 0.19  &    0.14 - 0.20  &        0.09      &    0.09 - 0.26 \\*
             &        0.48       &    0.11 - 0.33  &    0.19 - 0.28  &    0.21 - 0.30  &        0.13      &    0.13 - 0.38 \\* \midrule
$^{136}$Xe   &        0.46       &    0.17 - 0.74  &    0.28 - 0.39  &        0.19      &        0.12      &         -        \\*
             &        0.67       &    0.26 - 1.10  &    0.41 - 0.57  &        0.29      &        0.18      &         -        \\* \midrule
$^{150}$Nd   &          -      &        0.04      &         -       &    0.06 - 0.09  &        0.17      &    0.04 - 0.13 \\*
             &          -      &        0.06      &         -       &    0.09 - 0.13  &        0.25      &    0.05 - 0.18 \\* \bottomrule
\end{longtable}

\end{document}